\documentclass[a4paper,
               biblatex,     % biblatex is used
               hyphens,      % allow \url to hyphenate at "-" (hyphens)
               ]{jacow}
\usepackage{pdfpages,multirow,ragged2e} %
\makeatletter%
	\ifboolexpr{bool{xetex}}
	 {\renewcommand{\Gin@extensions}{.pdf,%
	                    .png,.jpg,.bmp,.pict,.tif,.psd,.mac,.sga,.tga,.gif,%
	                    .eps,.ps,%
	                    }}{}
\makeatother

\ifboolexpr{bool{xetex} or bool{luatex}} % test for XeTeX/LuaTeX
 {}                                      % input encoding is utf8 by default
 {\usepackage[utf8]{inputenc}}           % switch to utf8

\usepackage[USenglish]{babel}

\ifboolexpr{bool{jacowbiblatex}}%
 {%
  \addbibresource{ref.bib}
 }{}
\begin{document}

\title{Machine learning assisted Bayesian calibration of an accelerator digital twin from orbit response data\thanks{Work supported by Brookhaven Science Associates, LLC under Contract No. DE-SC0012704 with the U.S. Department of Energy and by DOE-NP No. DE SC-0024287.}}

\author{C. Kelly\textsuperscript{3,}\thanks{ckelly@bnl.gov},  K. A. Brown\textsuperscript{1}, G. H. Hoffstaetter\textsuperscript{1,2},  W. Lin\textsuperscript{1}, N. M. Urban\textsuperscript{4} \\
\textsuperscript{1}Collider-Accelerator Department, Brookhaven National Laboratory, Upton, NY, USA \\
\textsuperscript{2}CLASSE, Cornell University, Ithaca, NY, USA \\
\textsuperscript{3}Computational Science Department, Brookhaven National Laboratory, Upton, NY, USA \\
\textsuperscript{4}Applied Mathematics Department, Brookhaven National Laboratory, Upton, NY, USA}
	
\maketitle

\begin{abstract}
Digital twins of particle accelerators are used to plan and control operations and to design data collection campaigns. Accurate modeling typically requires knowledge of quantities that are hard to measure directly, e.g., magnet alignments, transfer functions relating power supply currents to magnetic fields, magnet nonlinearities, and stray fields. In this work we introduce multiplicative parameters to the quadrupole transfer functions to parametrize these effects. We use Bayesian methods to probabilistically estimate these parameters and their uncertainties by calibrating the Bmad digital twin to beam measurements performed at the AGS Booster at Brookhaven National Laboratory. The inference is computationally accelerated using a machine learning emulator of the physical accelerator digital twin trained to a perturbed-parameter ensemble of Bmad simulations. The result is a joint posterior distribution over the parameters constrained by the data, taking into account beam monitor errors. Incorporating estimates of the parameters into the digital twin is shown to result in a significant improvement in the quality of the model and provides error bars on the model parameters and predictions.

\end{abstract}

\section{Introduction}

The Booster of the Alternating Gradient Synchrotron (AGS) acts as the injector for the AGS, accelerating particles from 200 MeV to 1.5 GeV~\cite{bnlnews} for injection into the AGS. It also serves as a heavy-ion source for the NASA Space Radiation Laboratory (NSRL). As one of the most upstream components in the acceleration chain, accurate control of beam properties in the Booster was indispensable to providing high quality beams to the Relativistic Heavy Ion Collider (RHIC) and is required for the future Electron Ion Collider (EIC). 

An improved understanding of magnet properties in the Booster is essential for better beam control. A comparison of measured and simulated orbit response data reveals that the current Booster digital twin is inaccurate at the level of several percent. The overarching goal of our research is to utilize Bayesian uncertainty quantification (UQ) techniques to identify and model the sources of these discrepancies.

As a first step in this direction, we focus on the uncertainties associated with the 48 quadrupole magnets, as these are expected to play a dominant role. We introduce magnet-dependent parameters to the transfer functions that relate the magnetic field to the power supply current, allowing the unknown sources of error to individually affect the magnet's impact on the beam orbit. 

The values of these parameters are constrained using experimental data via Bayesian inference. The resulting posterior distribution provides data-constrained values for the parameters that are then incorporated into the digital twin to improve the modeling. 

\section{Orbit Response Measurement}

\begin{figure}[!htb]
    \centering
    \includegraphics[width=0.6\columnwidth]{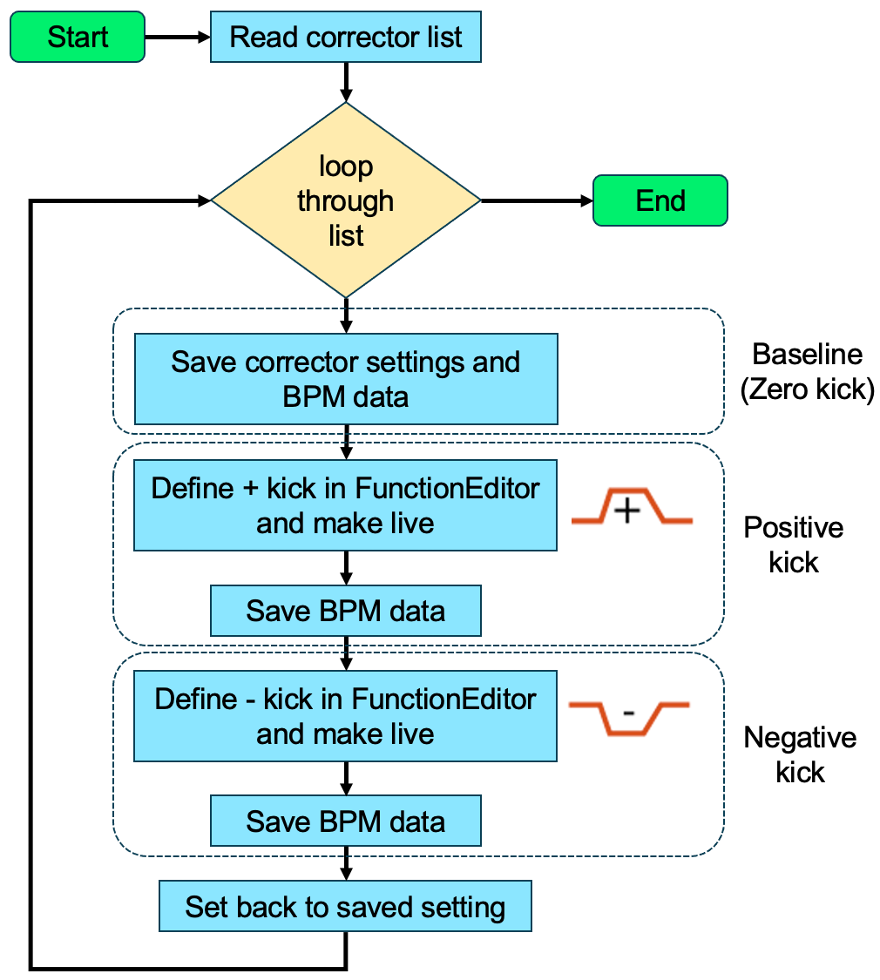}
    \caption{Work flow of orbit response measurement script.}
    \label{fig:script}
\end{figure}

 To collect suitable data from the Booster we developed a script~\cite{script} that changes the particle beam's closed orbit by setting each corrector magnet in turn to three settings: zero kick (baseline value), positive kick, and negative kick, while leaving all other correctors in their baseline setting. After setting the corrector, live beam position monitor (BPM) data and all the magnet settings are saved. The script work flow is outlined in Fig.~\ref{fig:script}.

The script was applied to both the 24 horizontal ($x$) and 24 vertical ($y$) correctors in the Booster. Each corrector was set to $\pm$22A between 50 and 110 milliseconds during the Booster magnet cycle. In this work we utilize only the data collected at a flat-top time of 92ms. For each setting, between 2 and 5 (typically 3) repeated measurements were taken over different cycles, for a total of 224 and 113 measurements for the horizontal and vertical correctors, respectively.

The data collected comprises both orbit data and magnet currents in the Booster, including dipoles, quadrupoles, sextupoles, and all correctors. These current values are obtained from hardware read-back monitors as opposed to the control room settings. These currents and control settings can be input to the physics simulation model, constructed in Bmad~\cite{Bmad}, to produce simulated orbit data. 

The Booster has 6 equivalently designed sections labeled "A" through "F". In each plane, the correctors, quadrupoles and BPMs are each grouped by the six ring segments, within each of which there are typically eight of each element, totaling 48. Within each segment, those elements for the horizontal plane are indexed as 2, 4, 6, and 8, and those in the vertical plane as 1, 3, 5, and 7. Due to hardware failures and machine geometry constraints, instead of 24 BPMs for each plane, only 15 BPMs are available for the horizontal  and 18 for the vertical. The complete list of valid BPMS is given in Tab.~\ref{tab:valid-bpms}

\begin{table*}[tb]

    \caption{A list of valid beam position monitors for each plane. \label{tab:valid-bpms}}
    
    \centering

\begin{tabular}{l|l}
\hline\hline
    Plane &  Valid BPMs \\
    \hline
    horizontal ($x$) & A6, A8, B4, B6, C2, C6, D2, D8, E2, E4, E6, E8, F2, F4, F8 \\
    vertical ($y$) & A1, A3, A5, A7, B1, B5, B7, C1, C3, C5, D3, D5, E5, E7, F1, F3, F5, F7
    \end{tabular}
\end{table*}

Data files of BPM readings are labeled by the corrector that was changed before the measurement. Data is missing for the horizontal-plane corrector D6, and for the vertical plane correctors A3-C5, leaving 23 and 14 choices of perturbed corrector for the horizontal and vertical planes, respectively. Furthermore, negative-kick data is missing for correctors A1, C6 and E7, and positive-kick data for E3. 

\begin{figure}[!htb]
    \centering
    \includegraphics[width=0.75\columnwidth]{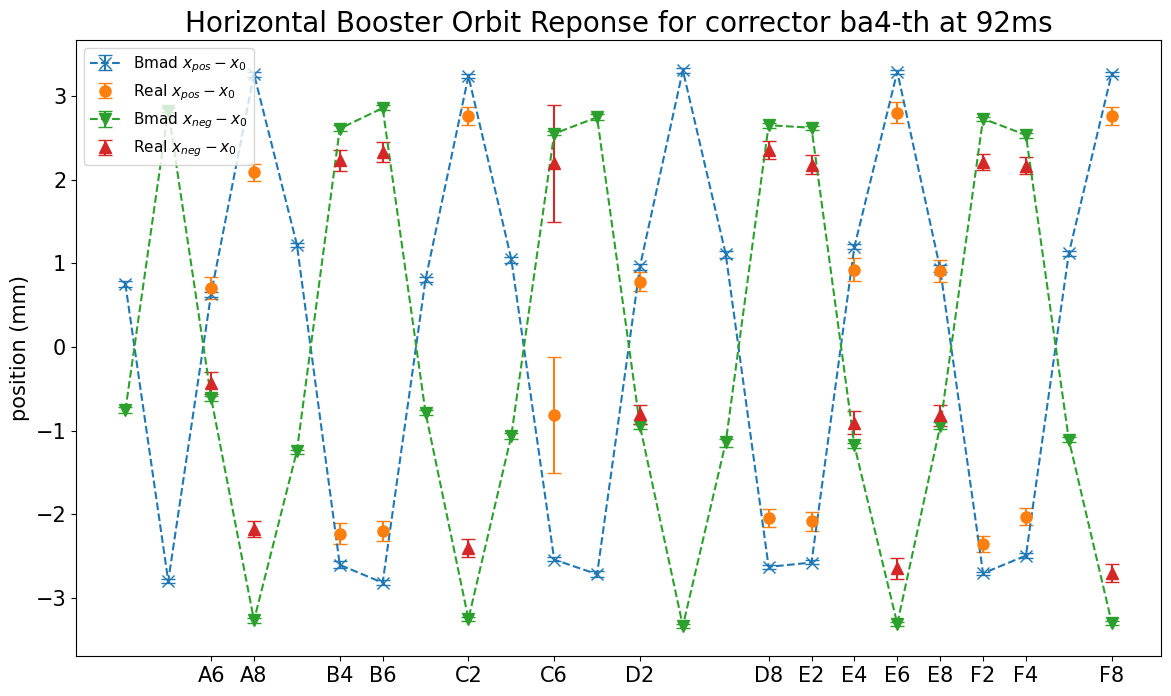}
    \caption{Comparison between measured and simulated horizontal orbit responses with corrector A4 set to $\pm 22$ A, data taken at 92 ms in the Booster cycle. The larger uncertainty for BPM C6 is due to a known hardware problem.}
    \label{fig:ba4_92ms}
\end{figure}

If the absolute location of all BPMs is accurately surveyed, it is possible to model the absolute beam orbit in the reference frame of the laboratory. However, for 100s of meters long accelerators, absolute BPM positions are often not known to sufficient accuracy. It is more relevant to the operation of the accelerator to instead consider orbit {\it responses} to changing the corrector settings, constructed from the difference between a base and perturbed BPM reading. (Using orbit responses also has the benefit of canceling some of the systematic uncertainties in the modeling.) While several correctors could be excited in various combinations, and the BPM readings could then be compared to simulated orbits, we focus on measurements in which a single corrector at a time is perturbed, for example, the orbit response of perturbing corrector A4 from its baseline setting of 0A to +22A with all other correctors remaining fixed at their baseline values (up to current fluctuations).

A comparison of the orbit differences between positive, zero, and negative corrector settings for corrector A4 is shown in Fig.~\ref{fig:ba4_92ms}. The error bars are estimated based on the fluctuations over the repeated measurements. We observe that the difference between measured and simulated differential orbits are mostly within 1 mm, but still much larger than the error bars of the measurements. This discrepancy implies there are sources of error in the real machine that are not included in the simulation model. 

As seen in Fig.~\ref{fig:ba4_92ms},  the BPM C6 has much larger measurement fluctuations due to known hardware issues. Prior to performing the inference we trimmed these and other outlier points from the data set using the procedure described in Appendix A.

\section{Booster Quadrupole Transfer Function}

There are 48 quadrupoles (24 for each plane) in the Booster, powered in series with the main bending dipoles. The simulation employs a fifth order polynomial to model the gradient of a quadrupole's magnetic field as a function of its power supply current, $I_q$:
\begin{equation}
    \frac{\partial B}{\partial r} = a_0 + a_1 \cdot I_q + a_2 \cdot I_q^2 + a_3 \cdot I_q^3 + a_4 \cdot I_q^4 + a_5 \cdot I_q^5
    \label{eq:b1}
\end{equation}
where $r$ is the radial distance from the center of the quadrupole in direction of its symmetry plane. The normalized gradient $k_1$ are:
%
%\begin{align}
 %   k_{1,H} &= \frac{1}{B \rho L_{H}} \left<\frac{\partial B_y}{\partial x}\right> \\
 %   k_{1,V} &= \frac{1}{B \rho L_{V}} \left<\frac{\partial B_x}{\partial y}\right>
%    \label{eq:k1_I}
%\end{align}
%
\begin{equation}
    k_{1,H} = \frac{1}{B \rho L_{H}} \left<\frac{\partial B}{\partial x}\right>,\hspace{0.5cm}
    k_{1,V} = \frac{1}{B \rho L_{V}} \left<\frac{\partial B}{\partial y}\right>
    \label{eq:k1_I}
\end{equation}
where $\frac{\partial B}{\partial x}$ and $\frac{\partial B}{\partial y}$ are obtained via Eq.~\ref{eq:b1} and $L_{H}$, $L_{V}$ are the lengths of magnets. The polynomial coefficients are derived from a least-squares linear regression fitting to match the tune measurement data from 1992 and 1993 ~\cite{tunecontrol}.

All quadrupoles receive the same current due to being wired in series, and the currents remain fixed in our data up to small power supply fluctuations. As such, the transfer function can be treated essentially as a magnet-independent constant. This provides a convenient location to introduce magnet-dependent properties that parametrize the model uncertainties. Specifically, we allow for magnet-dependent perturbations in the field strength through multiplicative parameters $var^{(x)}$ and $var^{(y)}$ for the $x-$ and $y-$plane quadrupoles, respectively. The resulting $k_1$ of the quadrupoles in the model are therefore defined as:
\begin{equation}
    (k_{1,H}')_i = var^{(x)}_i k_{1,H}, \hspace{0.3cm}
    (k_{1,V}')_i = var^{(y)}_i k_{1,V}
    \label{eq:k1_I_var}
\end{equation}
where $i$ labels the quadrupole magnet.

\section{Bayesian Uncertainty Quantification}

Uncertainty quantification is the process of identifying and quantifying uncertainties in models and simulations with the goal of exploring how uncertainties in model parameters or assumptions affect the outputs. A common method for UQ is Bayesian inference, which seeks to constrain or probabilistically ``fit'' model parameters using measurement data. Bayesian UQ takes expert knowledge of the model parameters $\theta$ as priors $p(\theta)$ as well as a probability model of the data-generating process $p(x|\theta)$ called the likelihood function. By constraining the parameters with data samples $\mathcal{D}_n = \{x_1, ..., x_n\}$, Bayesian UQ updates the model and calculates the posterior probability distribution $p(\theta|\mathcal{D}_n)$ of parameters $\theta$ given the data $\mathcal{D}_n$. According to Bayes' Theorem, the posterior distribution is given by~\cite{Wasserman2004}:
\begin{equation} P(\theta|x_1,...,x_n)=\frac{p(\mathcal{D}_n|\theta)p(\theta)}{p(\mathcal{D}_n)} = \frac{\mathcal{L}_n(\theta) p(\theta)}{c_n}
    \label{eq:Bayes}
\end{equation}
If the data points $\mathcal{D}_n = \{x_1, ..., x_n\}$ are statistically independent, then the likelihood function $\mathcal{L}_n(\theta) = \Pi_{i=1}^{n}p(x_i|\theta)$ factorizes into the product of likelihoods of each individual data point. The denominator of Eq.~\eqref{eq:Bayes}, $c_n = \int \mathcal{L}_n (\theta) p(\theta)\,d\theta$, is referred to as the evidence, which is analogous to the partition function in statistical mechanics and usually does not need to be calculated when using Monte Carlo sampling, as we will use here. In summary, the posterior is proportional to the product of the likelihood and the prior:
\begin{equation}
    p(\theta|\mathcal{D}_n) \propto \mathcal{L} (\theta) p(\theta)
\end{equation}
The likelihood function for this analysis is given by the probability model for the BPM readings $x = m(\theta;I_q) + \epsilon$ where $m(\cdot;\cdot)$ is the prediction of (a machine learning emulator of) the Bmad model as a function of its uncertain parameters $\theta$ and known input currents $I_q$, and $\epsilon$ is an \textit{iid} normal random error variable.

In our case, priors are required for the $24\times 2$ multiplicative parameters in the transfer functions (henceforth, "vars") described in the previous section. As this work is exploratory, little information is known regarding the expected values of the vars other than that their prior distributions should be peaked at the null value of unity (i.e., the most likely setting involves no multiplicative field strength correction to the assumed nominal values). We therefore employ lognormal priors, a conventional choice for multiplicative parameters that ensures no directional bias upwards or downwards on the log scale, and adjust the parameters such that the mean is fixed to unity and the variance to some chosen value that parametrizes our assumptions.

The likelihood function computes the expected orbit given specific values of the vars and of the "controls": the settings for the correctors and the associated dipole and vertical and horizontal quadrupole and sextupole currents. (For our data the sextupole currents are typically very small, ${\cal O}(0.05A)$, but are included regardless.) During inference, these control settings (baseline, positive, and negative) are obtained from the read-back currents stored in the files rather than the control room settings as these are a more reliable indicator of the actual currents. At present we do not attempt to model any uncertainties associated with this or of the current monitor hardware.

To map the controls to the orbit we employ a surrogate model of the Bmad simulator (see below), which provides expected beam positions at the locations of each of the 15 horizontal and 18 vertical BPMs, as a function of the controls and vars. To model the aleatoric errors in the BPMs we use the simulated values as the means of a normal distribution with a variance of $\epsilon=\sqrt{2}\sigma_{\rm BPM}$ where $\sigma_{\rm BPM}=0.15$ mm is the expected BPM error and the $\sqrt{2}$ is required due to the output being an orbit difference.

We perform the inference using the Julia "Turing" package~\cite{turing} to sample the posterior distribution.

\section{Surrogate Model}

Efficient sampling of the high-dimension posterior distribution is achieved using the "No U-turn Sampler", a variant of "Hamiltonian Monte Carlo" Markov chain algorithm~\cite{hoffman2011nouturnsampler}. This algorithm requires accurate evaluation of the derivatives of the simulated result, something that Bmad is presently unable to provide. A fast simulation is also important as the sampling can require many thousands of evaluations. These requirements spurred the introduction of a machine-learning surrogate model for the Bmad simulation.

The surrogate modeling is somewhat challenging due to the high dimensionality of the input features. There are $24\times 2$ corrector settings, 5 magnet currents and $24\times 2$ vars, totaling 101 input features. Likewise, there are 15+18 output features corresponding to the BPM orbit measurements. However, in practice we need only model data for which a single corrector in one of the two planes is perturbed from its baseline setting. Given that the influence of a $y-$plane corrector at near-zero current on the orbit in the $x-$plane (and vice versa) is negligible, we can ignore this cross-coupling and model the two orbit planes separately. The full set of both $x-$ and $y-$plane vars however, must still be included. Thus, we reduce the dimensionality of the model to 77 input features and 15 and 18 output features for the $x-$ and $y-$plane models, respectively.

We employ a ResNet-style architecture~\cite{he2015deepresiduallearning} with ${\sim}360k$ parameters comprising two hidden fully-connected layers of dimension $400\times 400$, each sandwiched within a skip connection, which allows each layer to model the residuals of the previous layers and stabilizes the models against vanishing parameter gradients. The models were trained using a large number of samples generated from direct Bmad simulations. We employed a Sobol' quasi-random scheme to sample the 77-dimensional input space, which has the benefit of providing a highly uniform distribution of sample points. The sampling ranges for the static magnet currents were chosen to encompass the full range of values encountered in the data. We include only training data in which a single corrector is set to $\pm22A$ with all other correctors set to $0A$, and allow for current fluctuations by sampling within a small range about these set points of sufficient width to encompass the range of fluctuations observed in the data. The data set includes an equal number of samples for each perturbed corrector to avoid bias. This approach allowed for a surrogate with much smaller validation losses within the data region than a model with the same architecture trained on a data set in which all correctors were allowed to vary uniformly across the entire ${\sim}44A$ value range.

Finding appropriate ranges for the vars was the biggest challenge for several reasons, the first of which was the lack of prior knowledge of an expected range. Ideally, we would choose a wide range of sample vars so as to not constrain our choice of priors. However, we found that for larger var ranges, a progressively larger fraction of the Bmad simulations failed due to orbit instability, and the models trained on the remaining samples became progressively less accurate even with hyperparameter tuning. These instabilities may be caused both by specific unfortunate configurations of the vars and also by resonances in the orbit "tune" -- the number of betatron oscillations per particle orbit -- which occur at half-integer values. To combat the resonance issues we subsampled our training data, keeping only those with a tune within 0.05 of the known values for the Booster (4.83 and 4.665 for the $x-$ and $y-$plane tunes, respectively). In order to obtain a distribution of tune values in the subsampled data set that peaks at the value of the baseline, unperturbed simulation -- which has a tune very close to the known Booster value -- it was necessary to sample the vars symmetrically about unity. If we instead sampled asymmetrically, e.g. preferring var values less than unity, the distribution of tunes in the subsampled data set would no longer be peaked near the true Booster tune and our surrogate model would therefore be trained on unrealistic data. 
%
%This constraint enforces a symmetric sampling distribution of var values about unity as this ensures the distribution of tunes peaks at the location of the baseline simulation, which has a tune very close to the known value. 

We then attempted to find an optimal var range that was both wide enough to not overly constrain our prior choices and also allowed for a sufficiently accurate surrogate that the model errors do not overly affect the inference. Our final choice is a var range of $0.94-1.06$.

With these choices of parameter range we generated around 2M samples for each plane, of which 820k ($x$) and 870k ($y$) samples remained after the tune cut. Retaining 5k samples for validation, we trained the model using the Julia "Flux" package, employing the Adam optimizer with a hand-tuned learning rate schedule and the mean-squared error (MSE) loss function, achieving final losses of $3.8\times 10^{-4}$ ($x$) and $1\times 10^{-4}$ ($y$) and no evidence of overfitting observed in the validation losses. To estimate the model error, we computed the standard deviation of the absolute difference between the Bmad simulation and the surrogate over all BPMs and validation samples, obtaining 0.057mm ($x$) and 0.017mm ($y$). While the error on the $x-$plane surrogate is larger, it is still $2.6\times$ smaller than the BPM error of 0.15mm. Nevertheless, for correctness we incorporated the model uncertainty into the inference by treating it as an aleatoric error combined in quadrature with the BPM errors.

%Separate models were trained for the $x-$ and $y-$plane orbits, each based on 820-890k simulated sample data points.

\section{UQ Results on Booster ORM}

\begin{figure}[t]
    \centering
    \includegraphics[width=0.7\columnwidth]{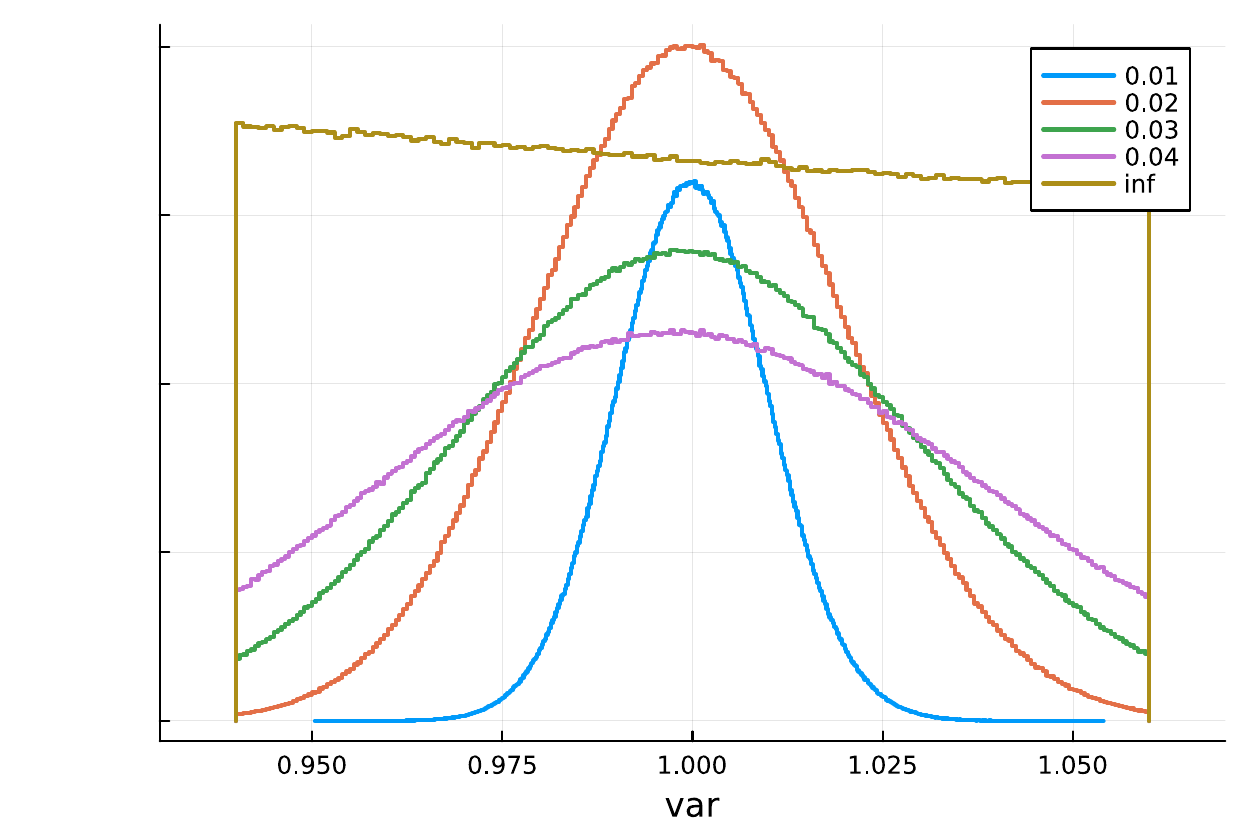}
    \caption{Histograms of the truncated lognormal priors for several choices of prior width.}
    \label{fig:prior_dists}
\end{figure}

As discussed above, we employ lognormal priors for the inferred parameters. Given that the deviations between the measurements and simulation are small, we will assume that the null result (${\rm var}_i=1$) is most likely and that discrepancies in the effective field strength of the magnets are on the order of a few percent. The prior distributions are therefore,
\begin{equation}
{\rm var}_i \sim {\rm Lognormal}(1 - \sigma^2/2, \sigma^2)
\end{equation}
for which
\begin{equation}
\overline{ {\rm var}_i } = 1,\ \ {\rm Var}({\rm var}_i)=\sigma^2 + {\cal O}(\sigma^4)\,,
\end{equation}
where we choose a uniform prior width, $\sigma$, for all quadrupoles. To account for the expected unreliability of the surrogate model beyond the range of the training data, we further truncate these distributions to the range 0.94--1.06. The effects of this truncation on the distributions is small for $\sigma < 0.04$ as illustrated in Fig.~\ref{fig:prior_dists}, and the effect on the mean can be discounted.

Each orbit response is described by an initial and final setting for the set of correctors. Although the data contain only measurements in which a single corrector is set to $\pm 22A$ with the others fixed at 0A (up to small fluctuations), there are typically three repeat measurements for each setting and we are free to independently vary which correctors are perturbed for the initial and final settings. As such, there are many possible orbit responses that can be incorporated into the inference. In principle, we could perform a single inference incorporating all possible responses; however, it is often useful to consider multiple inferences performed on different subsets of the data, allowing for an assessment of the robustness of the predictions. In this document we will employ only those orbit responses in which a single corrector at a time is perturbed from some initial setting $\alpha$ to a final setting $\beta$. We will label these data sets as $\alpha\to \beta$. 

We perform the inference using the No U-turn Sampler, requesting 2000 posterior samples. The initial values are chosen based on the maximum {\it a posteriori} (MAP) estimate. The Turing library automatically discards the first 1000 Markov steps for thermalization, and reports the effective sample sizes for each of the vars, which are observed to be typically ${\cal O}(1500)$ and only rarely as low as 700-900, thus ensuring sufficient samples for reliable uncertainty estimation. 

\subsection{Prior sensitivity analysis}

In order to identify an appropriate choice of prior width within the range specified above, we first perform a prior sensitivity analysis. We consider $\sigma \in \{0.01,0.02,0.03,0.04,\infty\}$, where the final two choices are included to assess the limiting behavior on the understanding that the distributions are heavily distorted by the truncation. For this analysis we will focus on a single $0A\to \pm22A$ subset, in which the $+22A$, $-22A$ and the two $0A$ measurements for each corrector are chosen randomly among the repeat measurements, while ensuring the two 0A measurements are distinct so as to reduce correlations between the input data samples. The results with $\sigma=\infty$ are obtained using the ${\rm Loguniform}(0.94,1.06)$ distribution, which represents a nearly-flat prior within our prior range (also shown in Fig.~\ref{fig:prior_dists}).

\begin{figure}[!htb]
    \centering
    \includegraphics[width=0.7\columnwidth]{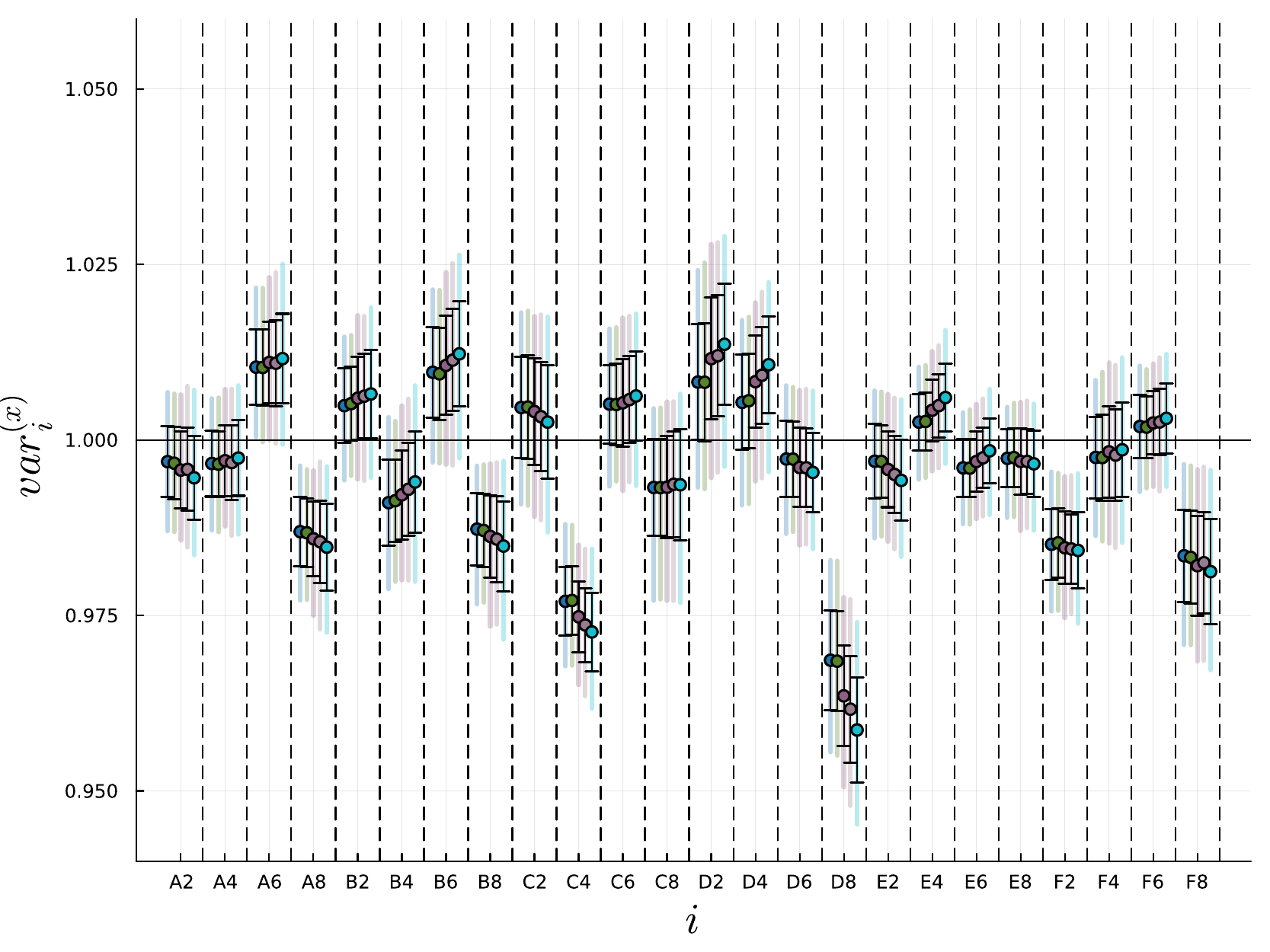} \\
    \includegraphics[width=0.7\columnwidth]{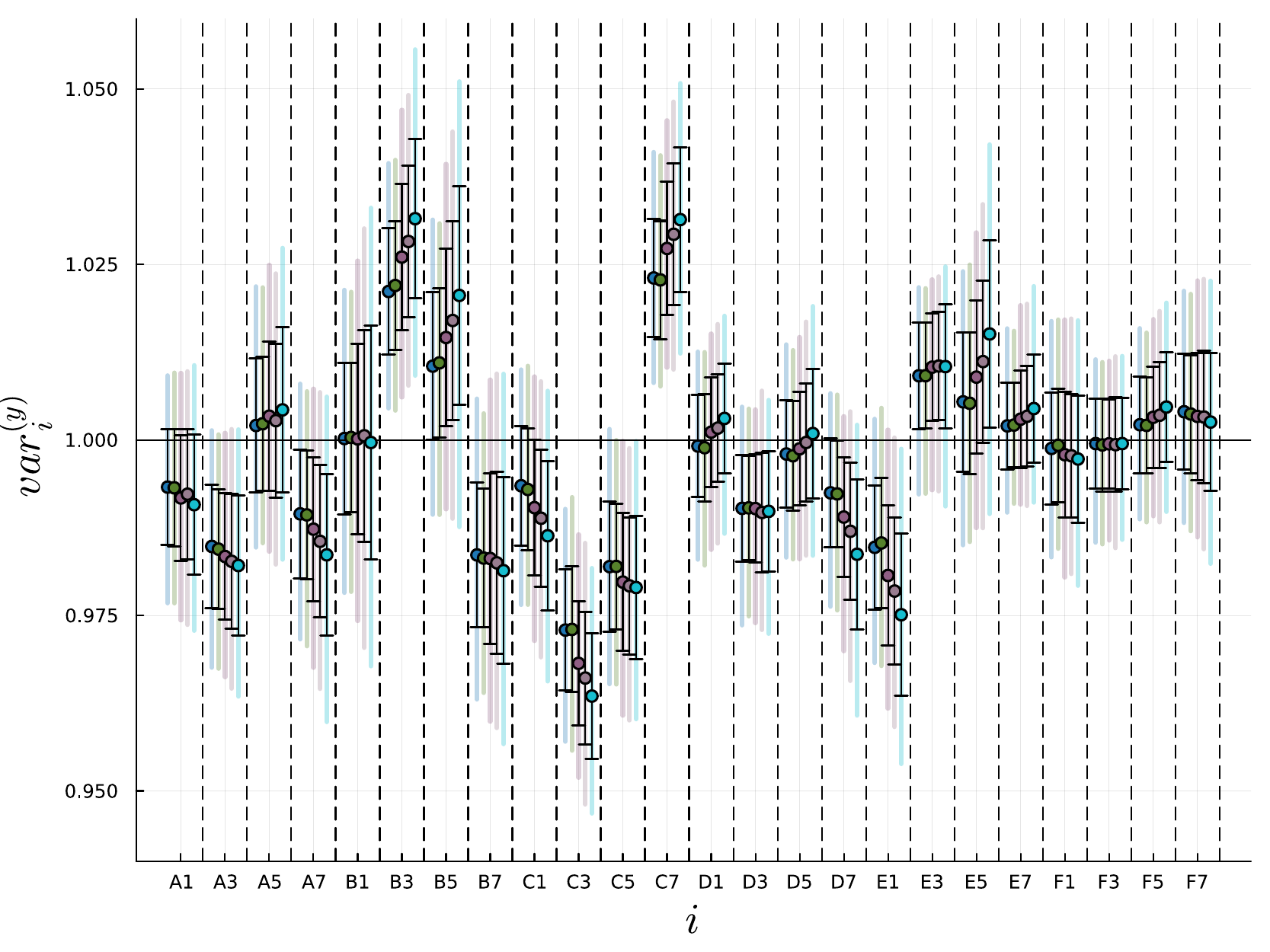} 
    \caption{The inferred $x-$ (upper) and $y-$plane (lower) vars for each of the 24 associated quadrupoles. For each magnet $i$ we plot left-to-right values obtained using prior widths $\sigma \in \{0.01,0.02,0.03,0.04,\infty\}$. The darker and lighter error bars indicate the 68\% and 95\% confidence bands, respectively.
    }
\label{fig:prior_stab}
\end{figure}

We plot the results of the inference for all vars in Fig.~\ref{fig:prior_stab}. We observe that the inferred values are typically stable as we increase the prior width, and the uncertainties do not grow substantially, suggesting the posterior is well constrained by the data. For some vars we observe a trend towards larger deviation from the null result of unity as the prior width is increased, although the results remain consistent within the uncertainties. Nevertheless, we will examine this effect in more detail, focusing on the $x$-plane var D8 which has the largest deviation from the null result.

\begin{figure}[!htb]
    \centering
    \includegraphics[width=0.5\linewidth]{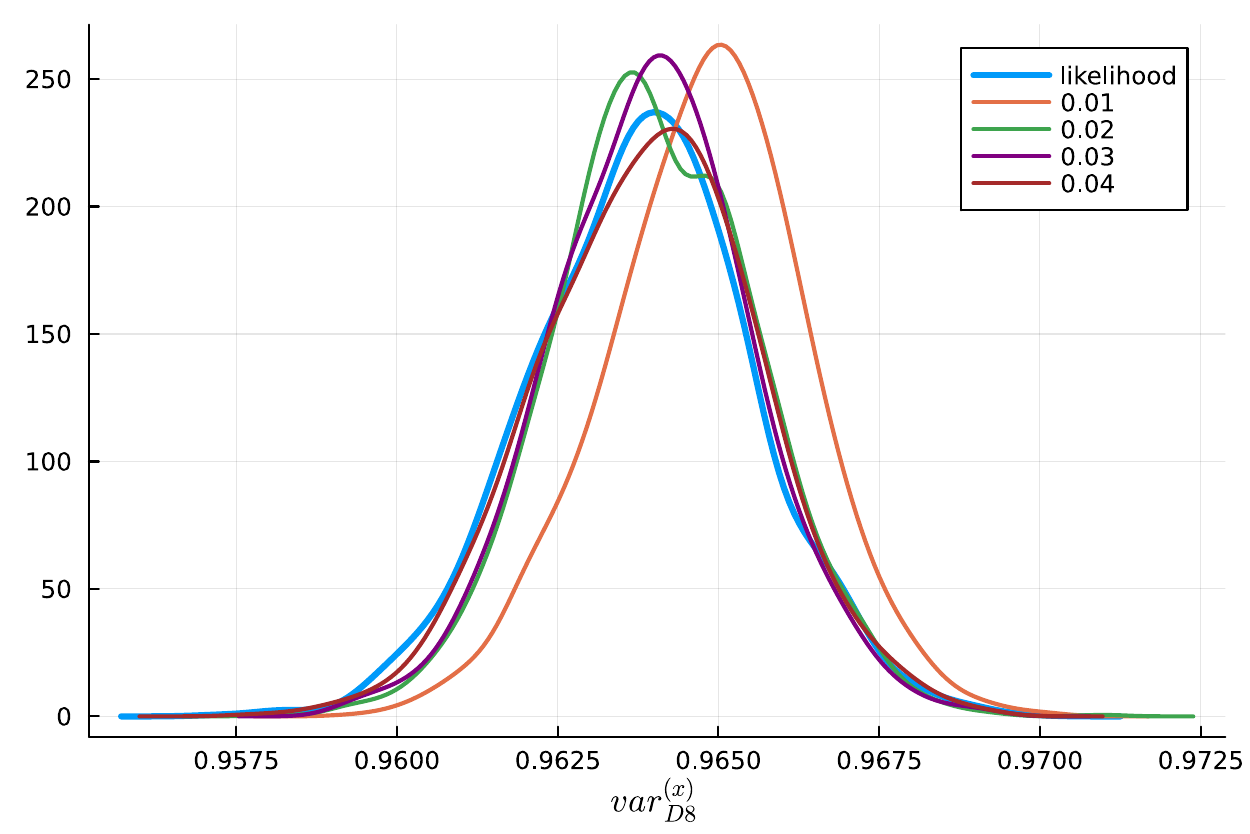}
    \caption{The result of performing inference on just the $x-$plane var D8. The blue curve shows the likelihood function, and the other curves show the posterior for several choices of prior width. }
    \label{fig:D8-inference-alone}
\end{figure}

To study var D8 in more detail, we repeat the inference but allow only D8 to vary, fixing all other vars to their best point estimates obtained from the above analysis with prior width $\sigma=0.03$. In doing so we isolate the effect of the data upon this parameter in the absence of potential cross-correlation between the variables. In Fig.~\ref{fig:D8-inference-alone} we plot the resulting likelihood function and posterior distributions for var D8. We observe that the likelihood function is strongly peaked at a value significantly below unity, indicating that the data are informative and that the parameter is not intrinsically prior-dominated. We also find only a very small prior-width dependence of the posterior in this analysis. 

\begin{figure}
    \centering
    \includegraphics[width=0.7\linewidth]{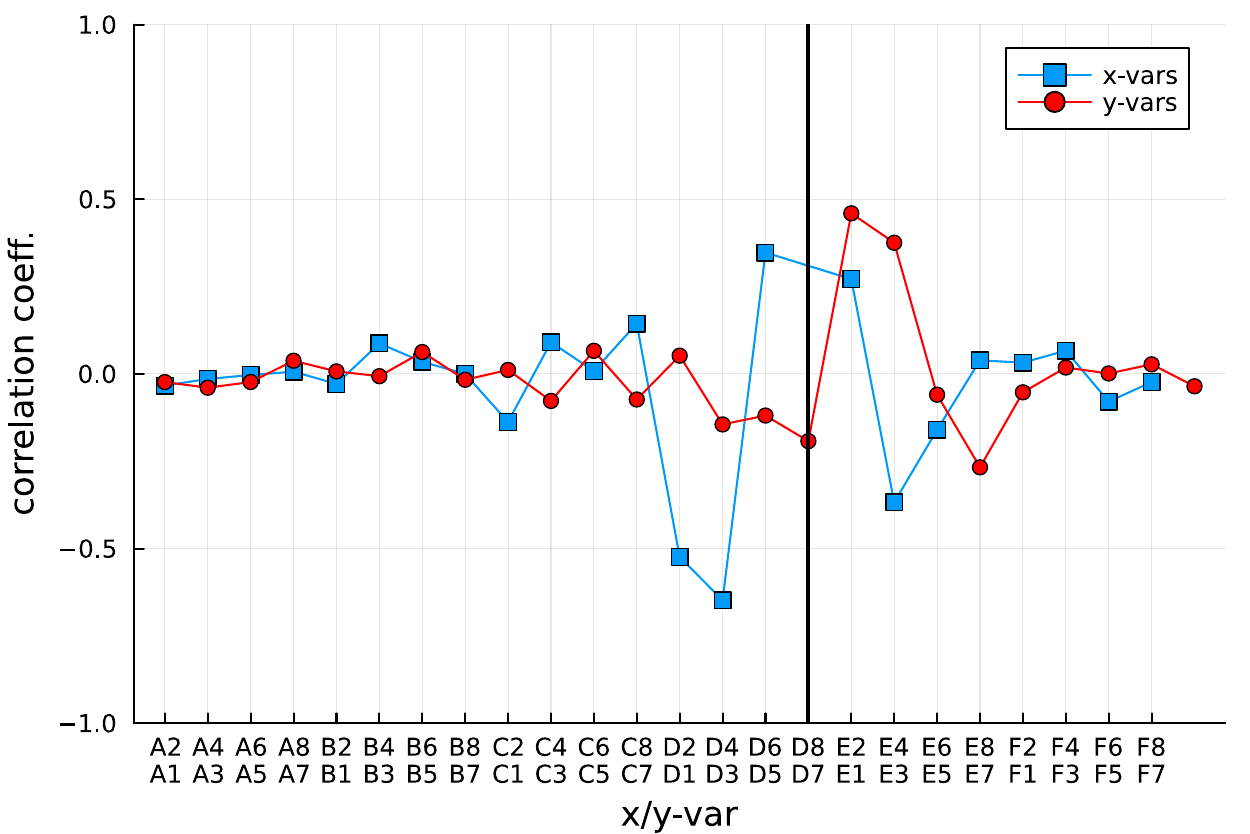}
    \caption{The correlation coefficient between var D8 and the other vars, obtained from the $0A\to \pm22A$ subset with prior width $\sigma=0.03$.}
    \label{fig:D8correlation}
\end{figure}

A likely explanation for the larger prior-width dependence of var D8 in our full inference is the development of a somewhat flatter direction in parameter space, where compensatory factors between different vars result in a reduced constraint of the data on the priors. Evidence for this conclusion can be found in the correlations between the posterior distributions of the vars. We plot this correlation for var D8 in Fig.~\ref{fig:D8correlation} for prior width $\sigma=0.03$, where we observe strong correlations between D8 and neighboring $x-$plane vars D2-E2, and with $y-$plane vars E1 and E3. Such correlation is expected given the physical proximity of the associated quadrupoles.

We emphasize that the role of the prior in Bayesian statistics is to parametrize our assumptions regarding the system. Dependence on the width or form of the prior is intrinsic to the method: only in the limit of infinite, perfectly informative data would the posterior be effectively independent of these assumptions. For finite data, the posterior distribution reflects a balance between prior beliefs and the likelihood supplied by the observations. A stronger prior-width dependence of a parameter simply indicates that the data alone are less informative about that parameter and that the prior plays a more influential role. This does not invalidate the inference; rather, it highlights the conditional nature of all Bayesian results, which must always be interpreted in the context of the specific assumptions encoded in the priors.

While we did not observe a strong prior-width dependence, we will nevertheless choose the largest (least informative) choice of prior that is not overly distorted by the truncation: $\sigma=0.03$, for the remainder of our analyses.

%TRY SHRINKING THE PRIOR FOR E1, DOES D8 STILL GO DOWN?

\subsection{Prediction stability}

\begin{figure}[!htb]
    \centering
    \includegraphics[width=0.7\columnwidth]{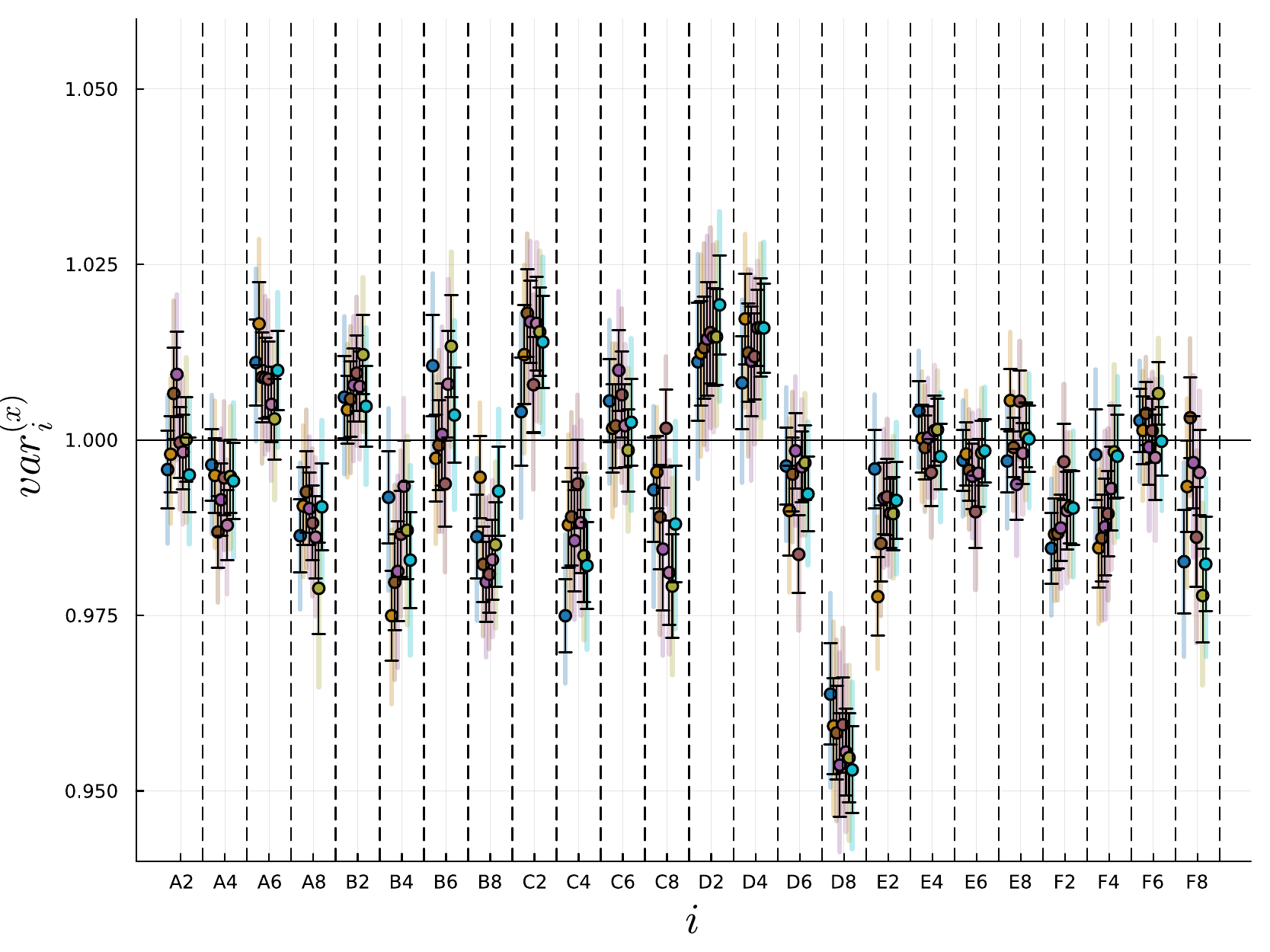} \\
    \includegraphics[width=0.7\columnwidth]{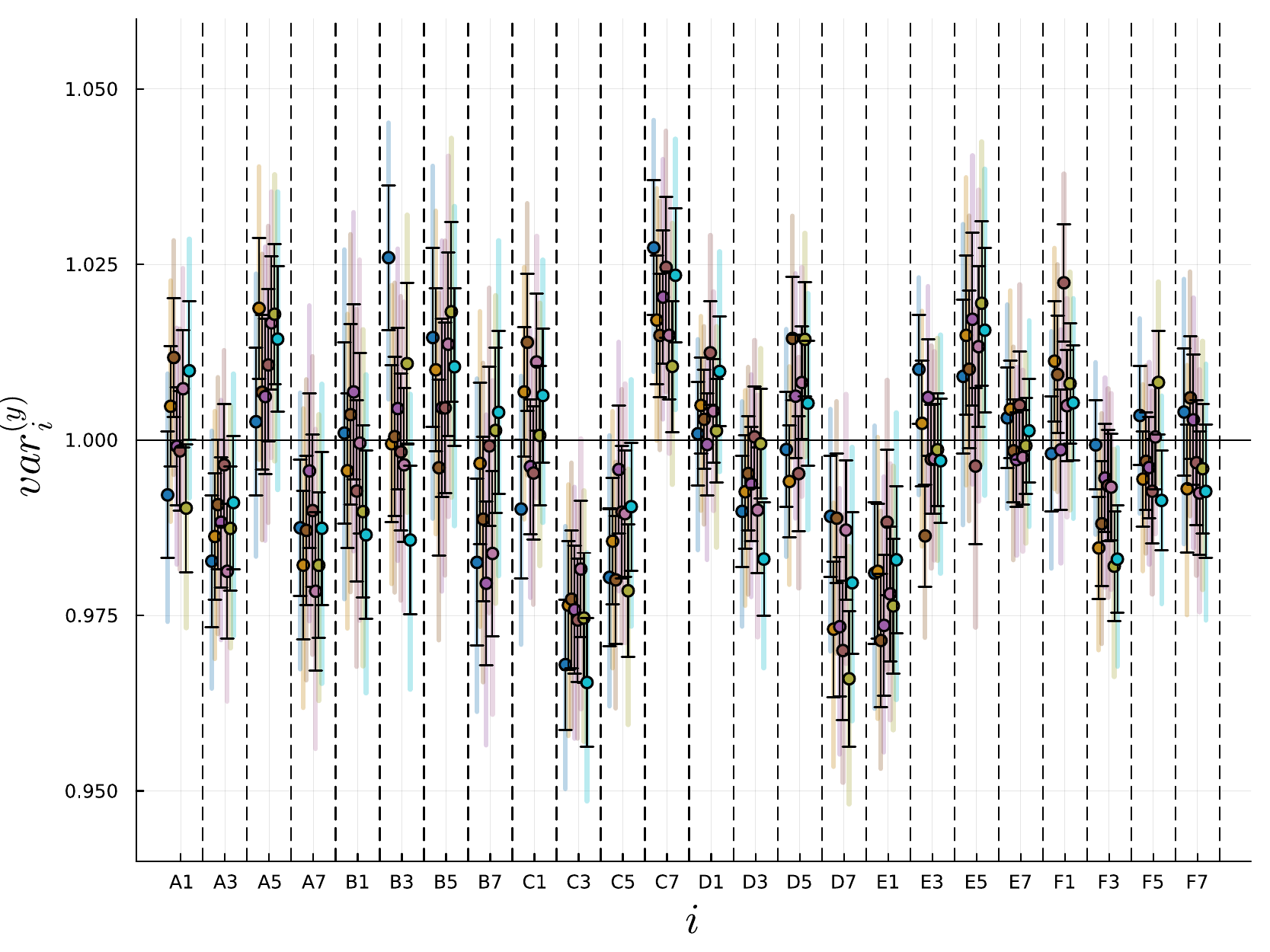} 
    \caption{The inferred $x-$ (upper) and $y-$plane (lower) vars for each of the 24 associated quadrupoles. For each magnet $i$ we plot left-to-right values obtained using eight different $0A\to \pm22A$ data subsets for which the  measurement for each perturbed/unperturbed corrector is chosen randomly among the available repeat measurements. The darker and lighter error bars indicate the 68\% and 95\% confidence bands, respectively.
    }
\label{fig:var_stab_pm22m0}
\end{figure}

In order to test the stability of our predictions, we generated seven additional $0A\to \pm22A$ datasets using the random selection mechanism described above. The inferred vars are shown in Fig.~\ref{fig:var_stab_pm22m0}, where we observe excellent consistency across all subsets. The largest variations typically correspond to those that showed larger prior sensitivity in our above analysis; such a sensitivity to small changes in the data can be expected if they are associated with flatter directions in parameter space. However, we note that quadrupole D8 remains consistently lower than unity in all cases.

\begin{figure*}[tb]
    \centering
    \includegraphics[width=0.65\columnwidth]{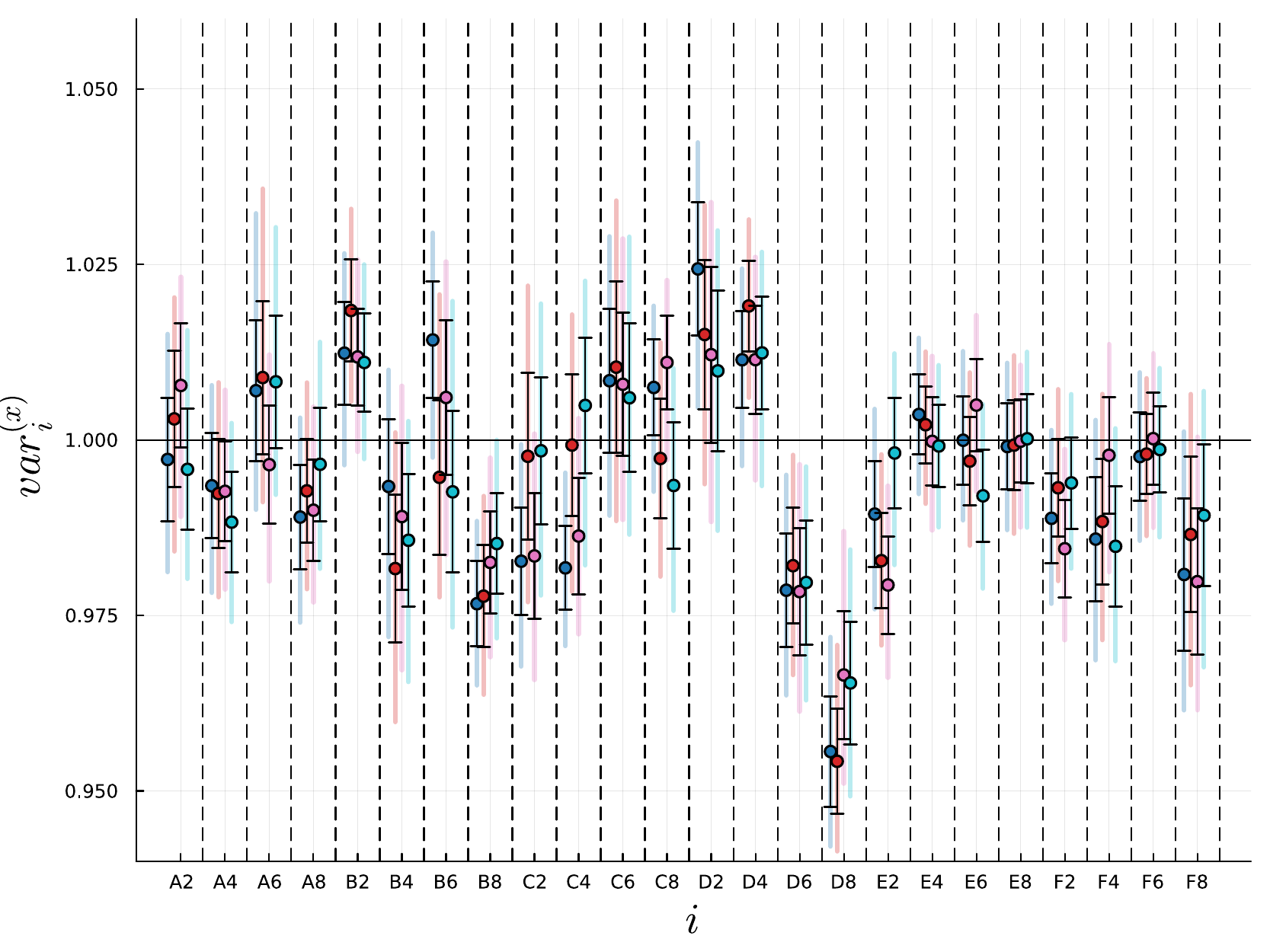}    
    \includegraphics[width=0.65\columnwidth]{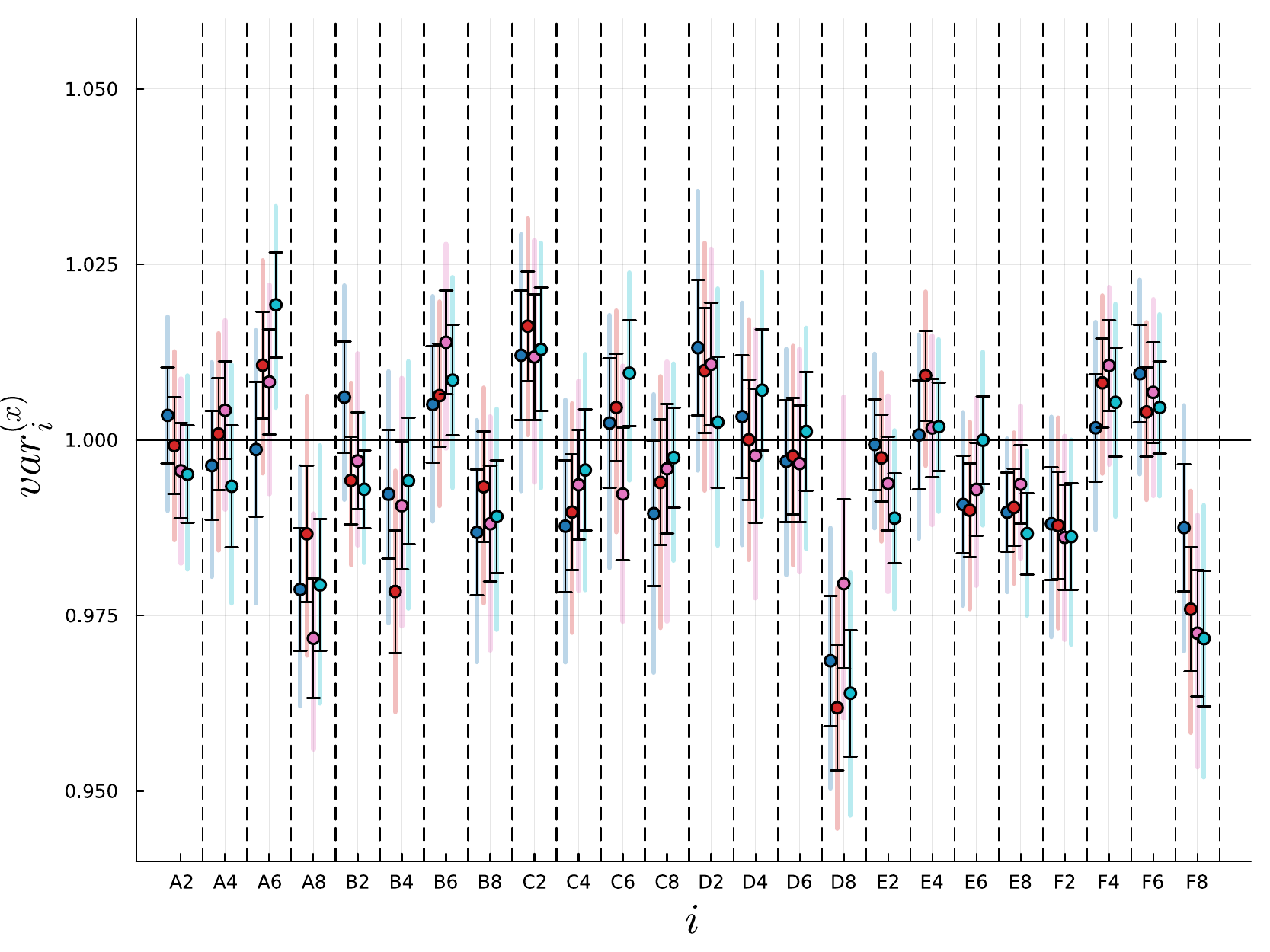}        
    \includegraphics[width=0.65\columnwidth]{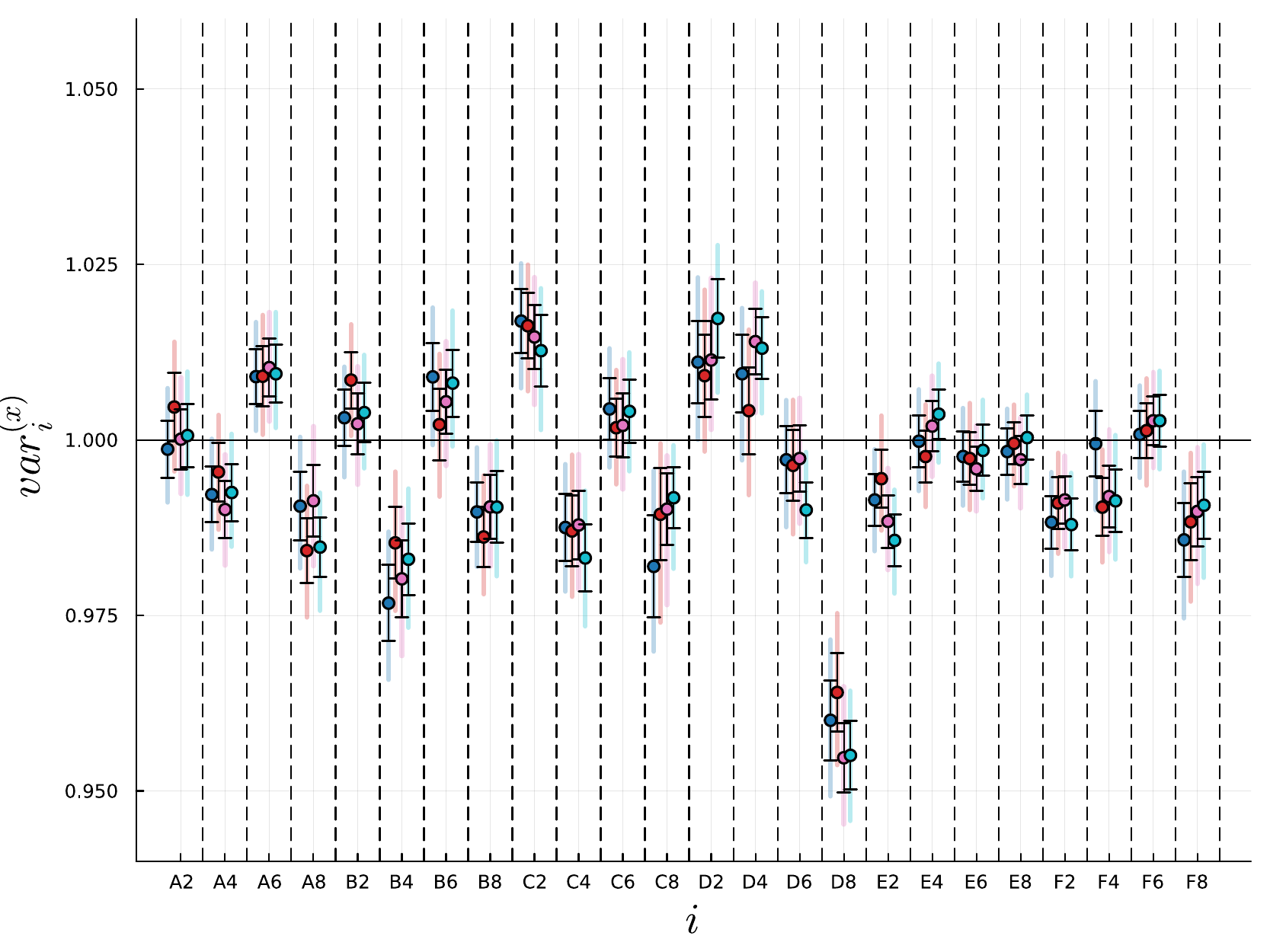}     \\
    \includegraphics[width=0.65\columnwidth]{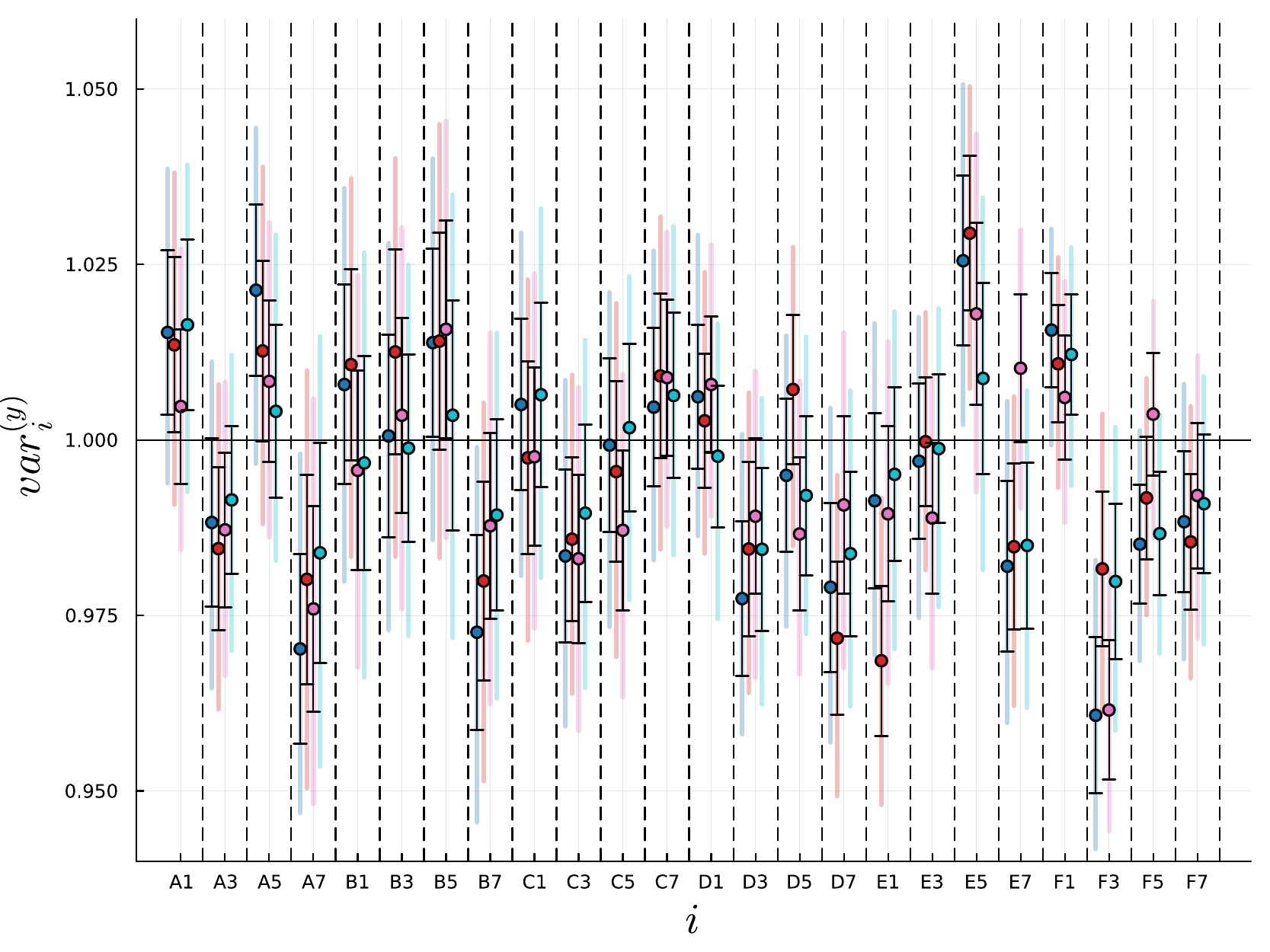}    
    \includegraphics[width=0.65\columnwidth]{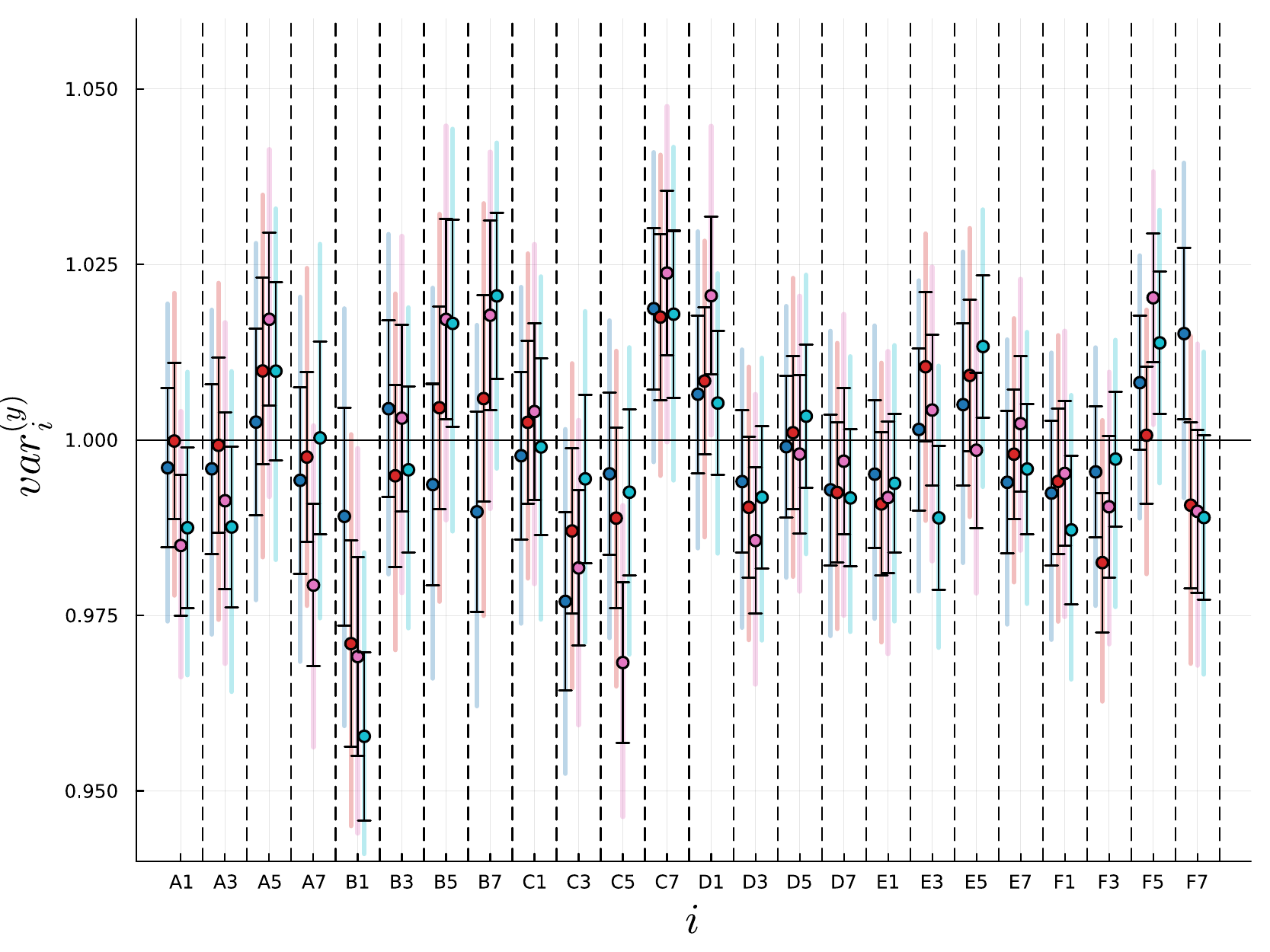}        
    \includegraphics[width=0.65\columnwidth]{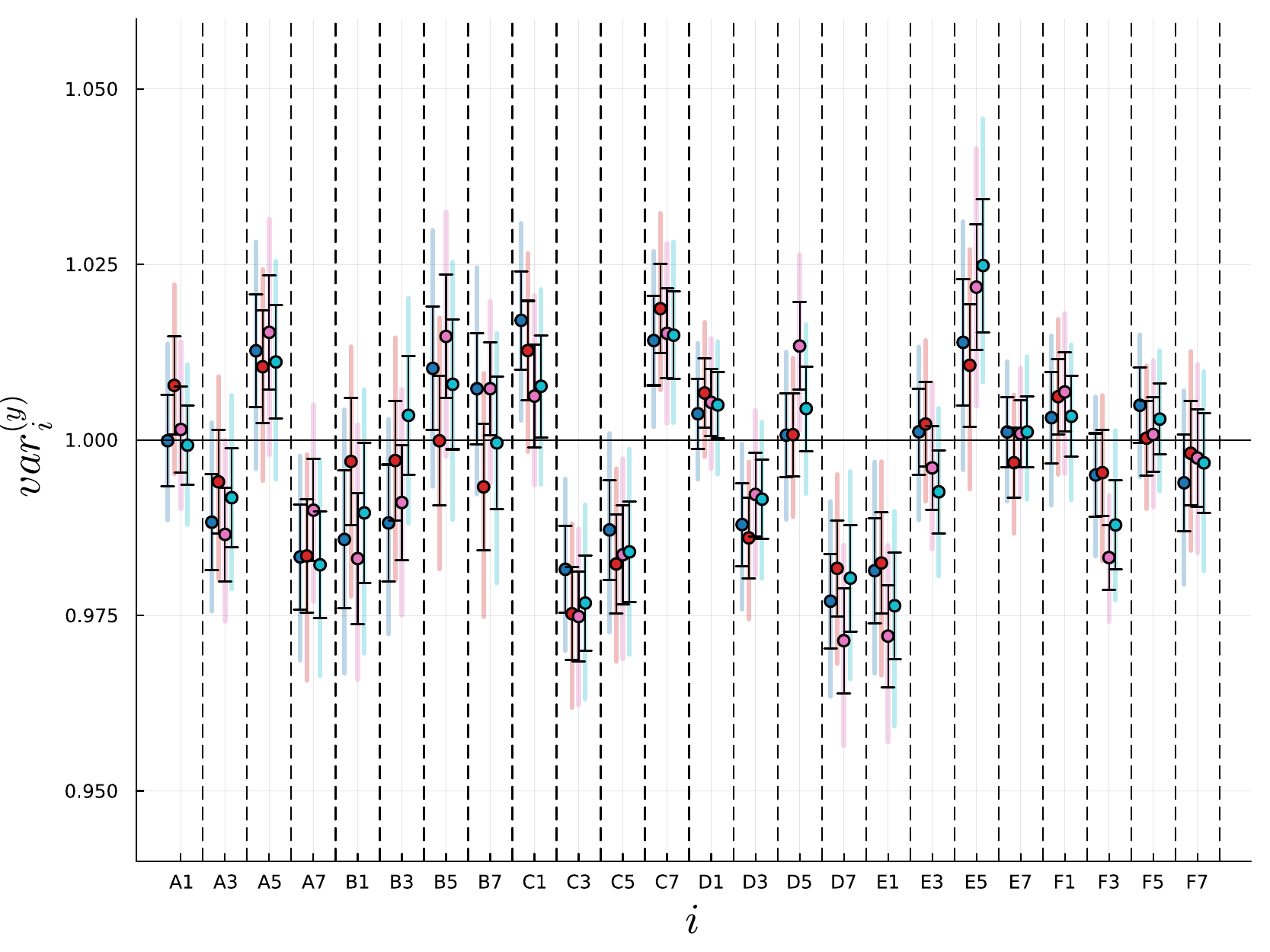}     \\
    \caption{The inferred $x-$ (upper) and $y-$plane (lower) vars for each of the 24 associated quadrupoles. The figures on each line correspond to the $0A\to +22A$, $0A\to -22A$, and $-22A \to +22A$ datasets, respectively. Within each figure we plot left-to-right the values for each magnet $i$ obtained using four different subsets for which the measurement for each corrector setting is chosen randomly among the available repeat measurements. The darker and lighter error bars indicate the 68\% and 95\% confidence bands, respectively.
    }
\label{fig:var_stab_other}
\end{figure*}

\begin{figure}[!htb]
    \centering
    \includegraphics[width=0.7\columnwidth]{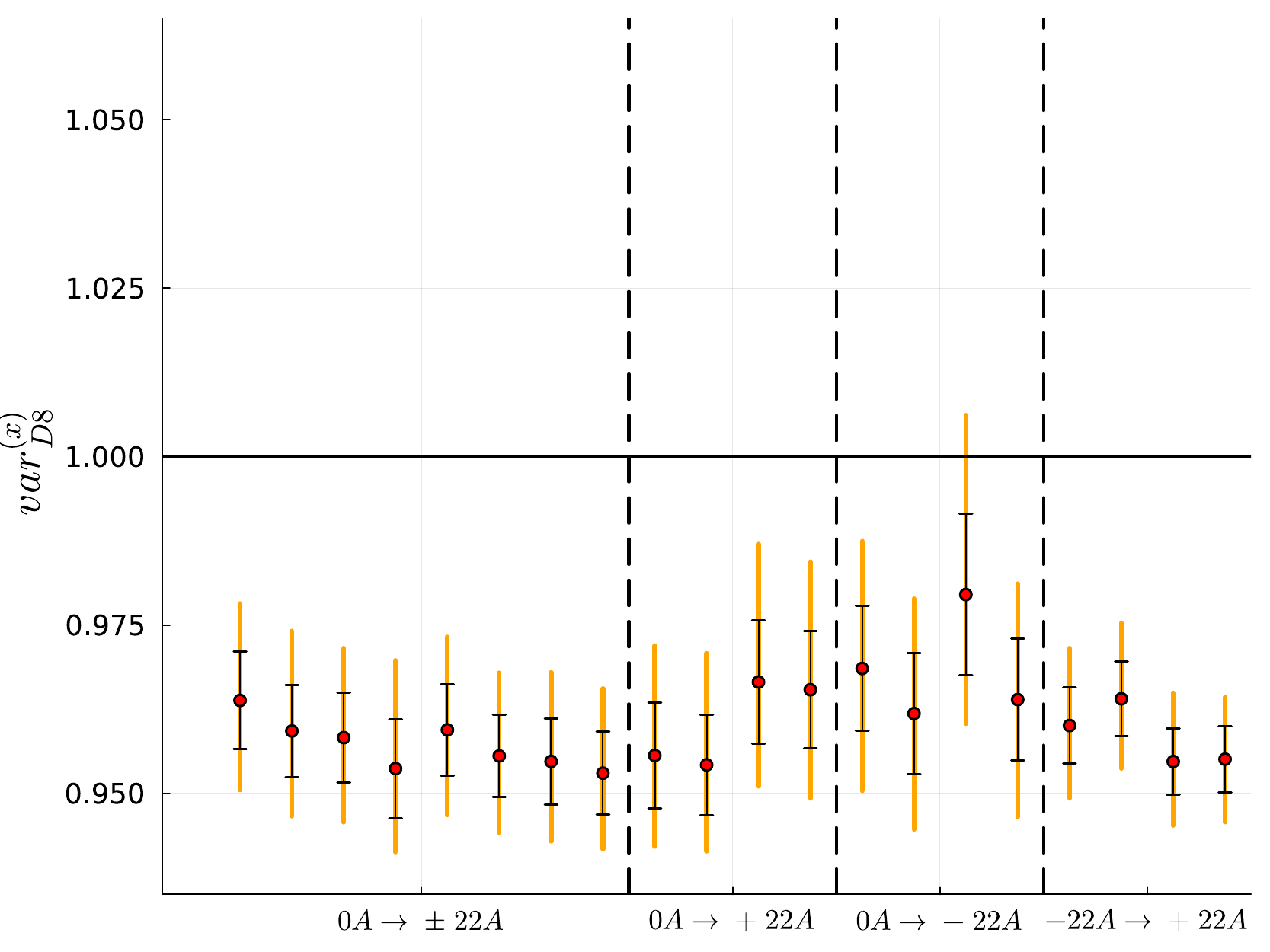}
    \caption{The inferred values for the $x-$plane var D8 over all choices of dataset explored within this document. The darker and lighter error bars indicate the 68\% and 95\% confidence bands, respectively.
    }
\label{fig:var_stab_D8}
\end{figure}

We also examined three alternative datasets: $0A\to +22A$, $0A\to -22A$, and $-22A \to +22A$. Here the first two are subsets of the $0A\to \pm22A$ dataset used previously, that include either the $+22A$ or $-22A$ set point in the inference but not both. For each dataset we again select randomly among the repeat measurements to generate four subsets. The inference results are plotted for each dataset separately in Fig.~\ref{fig:var_stab_other}, and all of our results for the D8 quadrupole are compared in Fig.~\ref{fig:var_stab_D8}. We observe strong consistency between the results. We also find that the results for the $-22A \to +22A$ typically have significantly smaller uncertainty, likely because the digital twin errors are amplified by combining two extreme corrector settings, thus making the data more informative.

\subsection{Final inference results}

\begin{table}[tbp]
\centering
\begin{tabular}{c|c||c|c}
\hline\hline
$i$ & ${\rm var}^{(x)}_i$ & $j$ & ${\rm var}^{(y)}_j$ \\
\hline 
A2 & 1.001(5)(20) & A1 & $1.002(8)(30)$ \\ 
 A4 & 0.993(5)(20) & A3 & $0.990(8)(30)$ \\ 
 A6 & 1.009(5)(20) & A5 & $1.012(9)(30)$ \\ 
 A8 & 0.988(6)(20) & A7 & $0.985(9)(30)$ \\ 
 B2 & 1.004(5)(20) & B1 & $0.99(2)(3)$ \\ 
 B4 & 0.981(7)(20) & B3 & $0.99(2)(3)$ \\ 
 B6 & 1.006(6)(20) & B5 & $1.01(2)(4)$ \\ 
 B8 & 0.989(5)(20) & B7 & $1.00(2)(3)$ \\ 
 C2 & 1.015(5)(20) & C1 & $1.011(9)(30)$ \\ 
 C4 & 0.986(6)(20) & C3 & $0.977(8)(20)$ \\ 
 C6 & 1.003(5)(20) & C5 & $0.984(8)(30)$ \\ 
 C8 & 0.988(8)(20) & C7 & $1.016(7)(20)$ \\ 
 D2 & 1.012(7)(20) & D1 & $1.005(5)(20)$ \\ 
 D4 & 1.010(7)(20) & D3 & $0.989(7)(20)$ \\ 
 D6 & 0.995(6)(20) & D5 & $1.005(8)(30)$ \\ 
 D8 & 0.958(7)(20) & D7 & $0.978(9)(30)$ \\ 
 E2 & 0.990(6)(20) & E1 & $0.978(9)(30)$ \\ 
 E4 & 1.001(5)(20) & E3 & $0.998(8)(20)$ \\ 
 E6 & 0.997(4)(20) & E5 & $1.02(2)(4)$ \\ 
 E8 & 0.999(4)(10) & E7 & $1.000(6)(20)$ \\ 
 F2 & 0.990(4)(20) & F1 & $1.005(7)(20)$ \\ 
 F4 & 0.993(6)(20) & F3 & $0.990(8)(20)$ \\ 
 F6 & 1.002(4)(20) & F5 & $1.002(6)(20)$ \\ 
 F8 & 0.989(6)(20) & F7 & $0.997(8)(30)$ \\ 
\end{tabular}
\caption{The final inferred var values. The left half of the table contains the $x$-plane vars, and the right half the $y$-plane vars. The errors in parentheses represent the 68\% and 95\% confidence bands, respectively. \label{tab-final-vars-all}}
\end{table}

\begin{figure}[!htb]
    \centering
    \includegraphics[width=0.7\columnwidth]{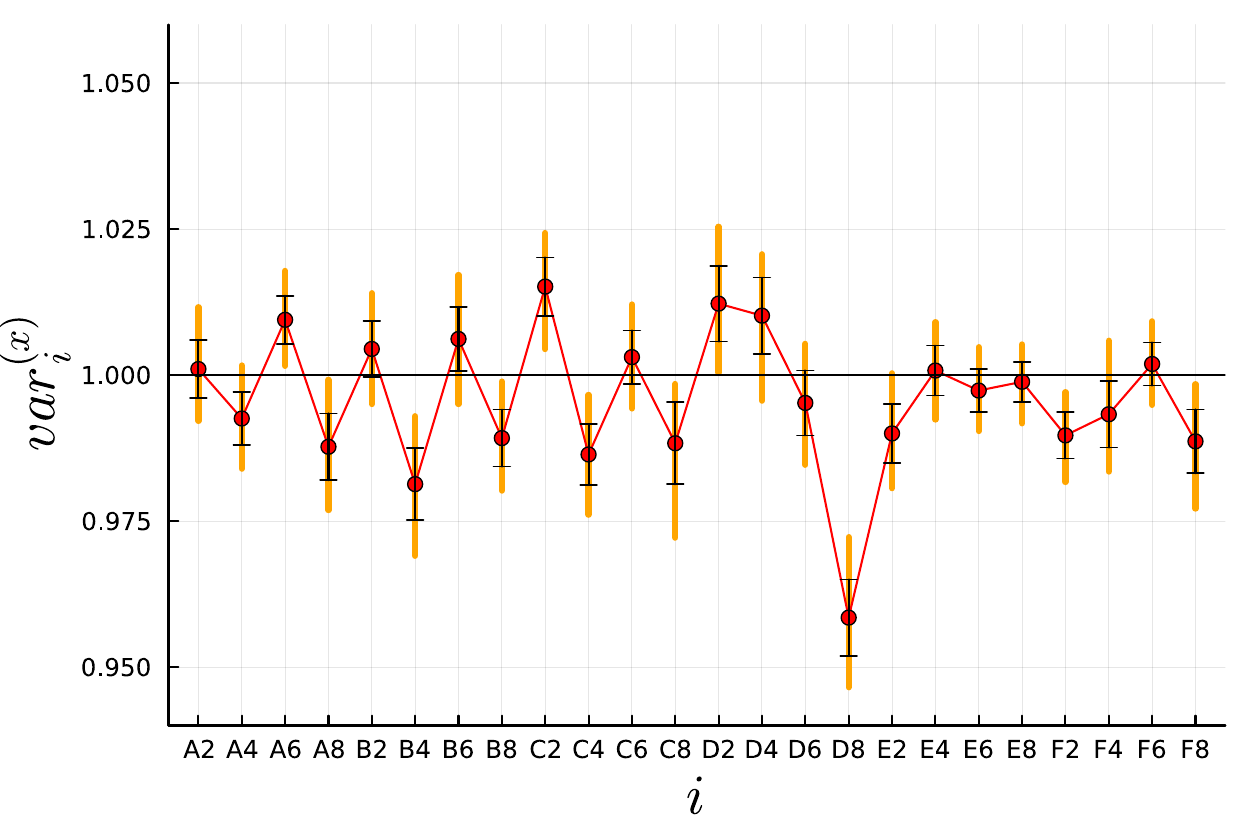} \\
    \includegraphics[width=0.7\columnwidth]{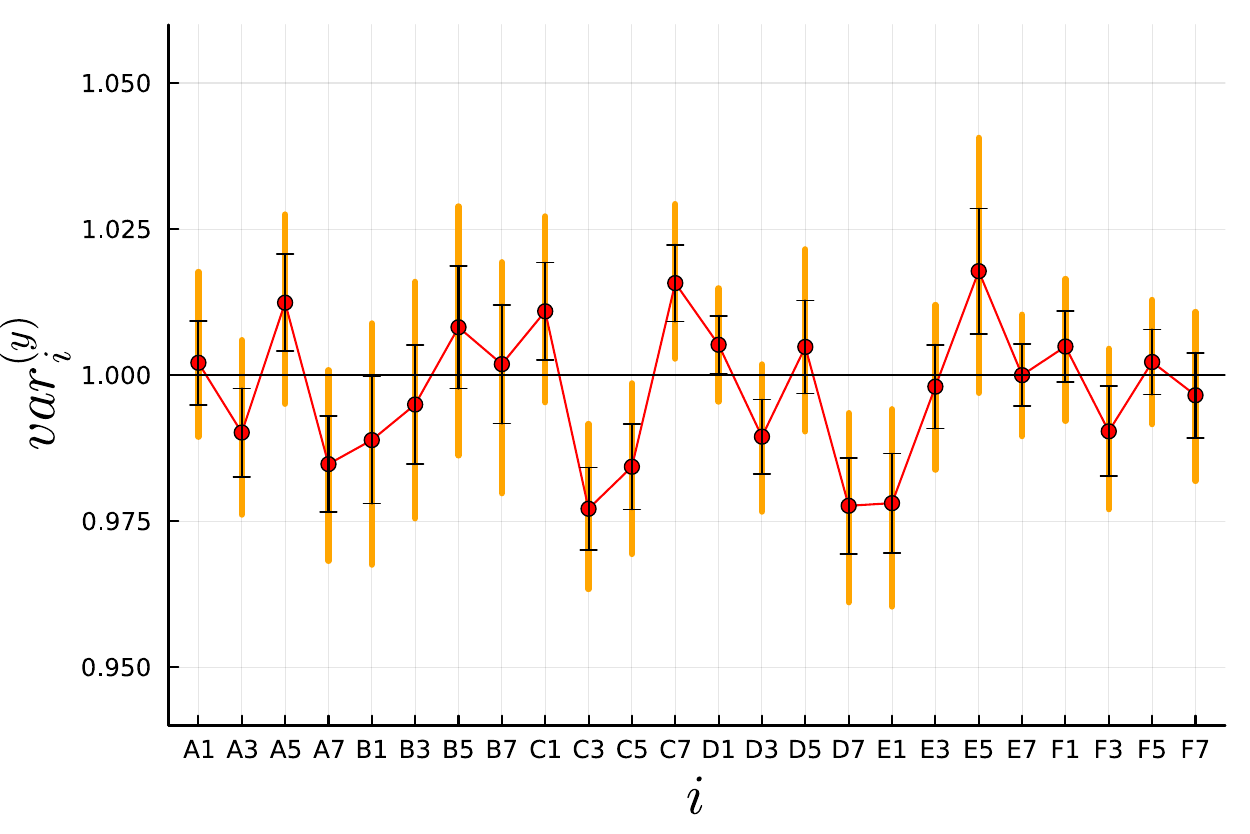} 
    \caption{The final posterior results for the inferred $x-$ (upper) and $y-$plane (lower) vars for each of the 24 associated quadrupoles. The darker and lighter error bars indicate the 68\% and 95\% confidence bands, respectively.}
    
\label{fig:vars_final}
\end{figure}

For our final results we focus upon the $-22A \to +22A$ combination that was found to produce the smallest uncertainty on our predictions. The posterior samples of the four random subsets are pooled to better account for the variation in the data. The final results for the inferred vars are shown in Fig.~\ref{fig:vars_final} and tabulated in Tab.~\ref{tab-final-vars-all}. We find that the majority of the results cluster around the null result of unity to within 2.5\%, although the deviations are often well outside of the 68\% confidence interval and are therefore statistically significant. There are also several vars for which the deviation from unity is outside of our 95\% confidence intervals, although only by a small amount.

\begin{figure}[tb]
\centering
\includegraphics[width=0.95\columnwidth]{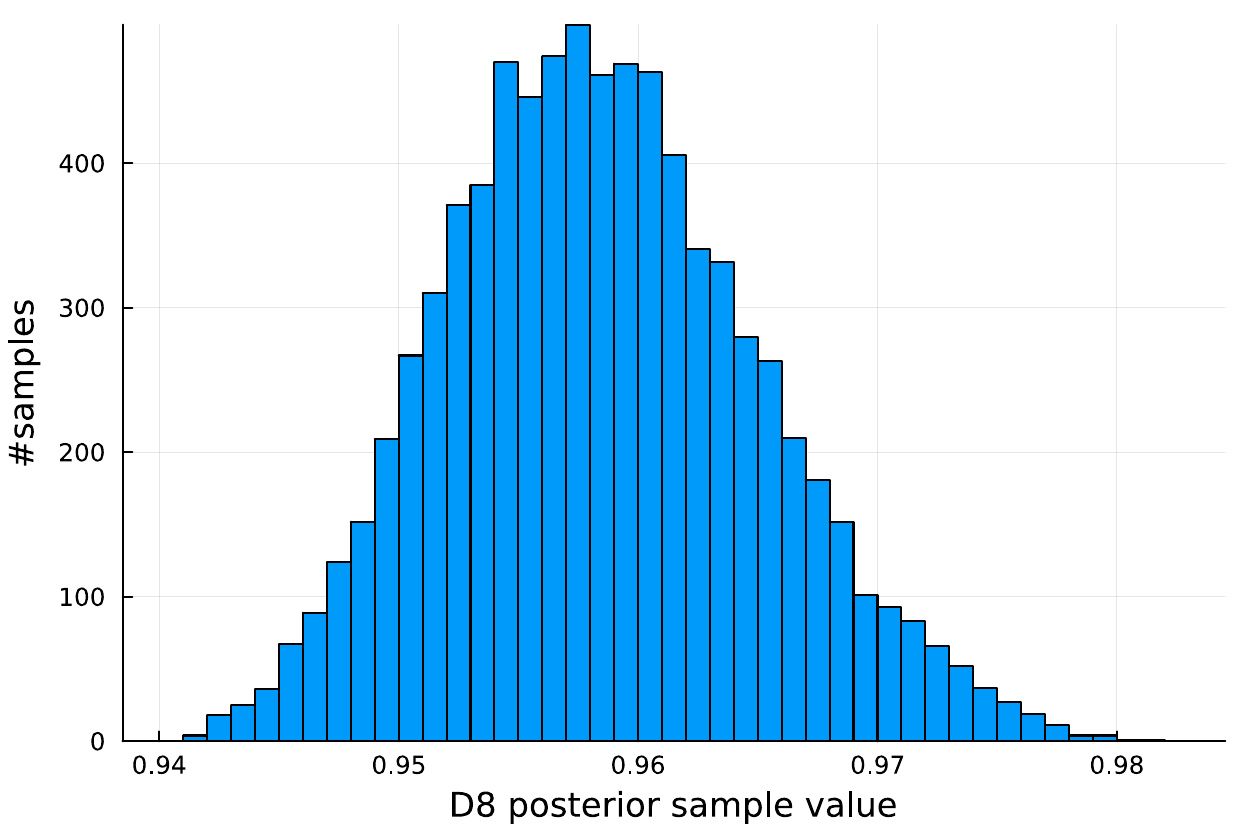}
\caption{A histogram of the posterior distribution of the D8 var. \label{fig-D8-posterior-hist}}
\end{figure}

The most significant deviation from the null result by far is for the $x-$plane quadrupole D8 with value $0.958(7)(20)$ (the parentheses indicating the $68\%$ and $95\%$ confidence bands, respectively), differing by nearly 5\% and far outside of our confidence intervals. This var is one that exhibited a larger prior-width dependence, although we note that even priors much more strongly peaked around the null result (cf. Fig.~\ref{fig:prior_stab}) gave values consistent with this final result. One point of concern may be that this value is close to the lower bound 0.94 of our surrogate model training data and the corresponding truncation of the prior distribution. To address this we plot a histogram of the posterior distribution for the D8 var in Fig.~\ref{fig-D8-posterior-hist}, where we observe little evidence of any resulting truncation of the posterior distribution. This strongly suggests that quadrupole D8 is actually behaving unexpectedly, although it remains the subject of further research to identify exactly what is driving this behavior.

%We emphasize again that this result should be interpreted in the context of the assumptions as encoded by our priors. We also note that the result is very close to the lower bound of the training data used for the surrogate model and the corresponding truncation of the prior distribution. However, it is clear that the data favors a lower value, as evidenced by Fig.~\ref{fig:D8-inference-alone}, the fact that the value dropped as the prior width was increased, and the general stability of the prediction under variations in the data. 

\begin{figure}[!htb]
    \centering
    \includegraphics[width=0.7\columnwidth]{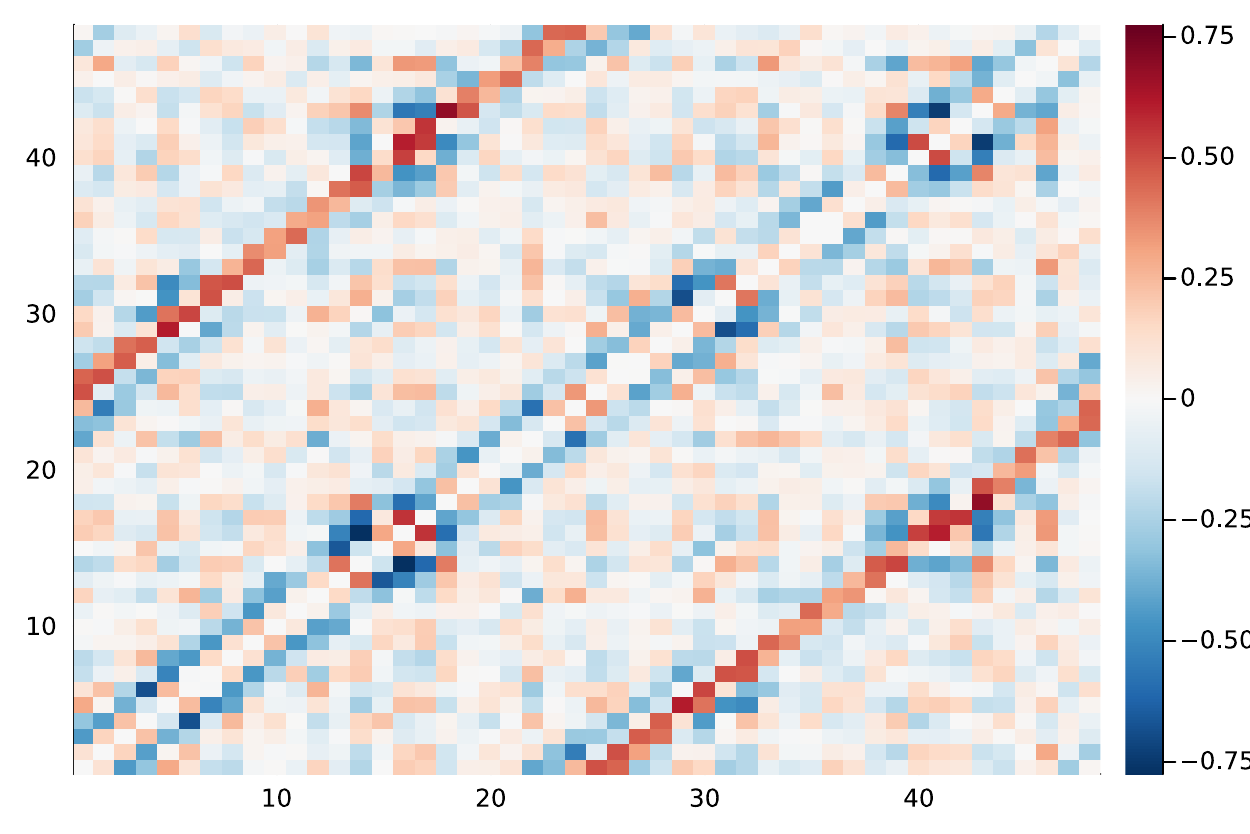}    
    \caption{A heatmap plot of the correlation matrix for the pooled posterior distributions of the vars after subtracting the diagonal unit matrix to suppress the self-correlations. Here, the $x-$plane quadrupoles are assigned indices $1-24$ and the $y-$plane quadrupoles $25-48$. }
    
\label{fig:cormat_final}
\end{figure}

\begin{figure}
    \centering
    \includegraphics[width=0.7\linewidth]{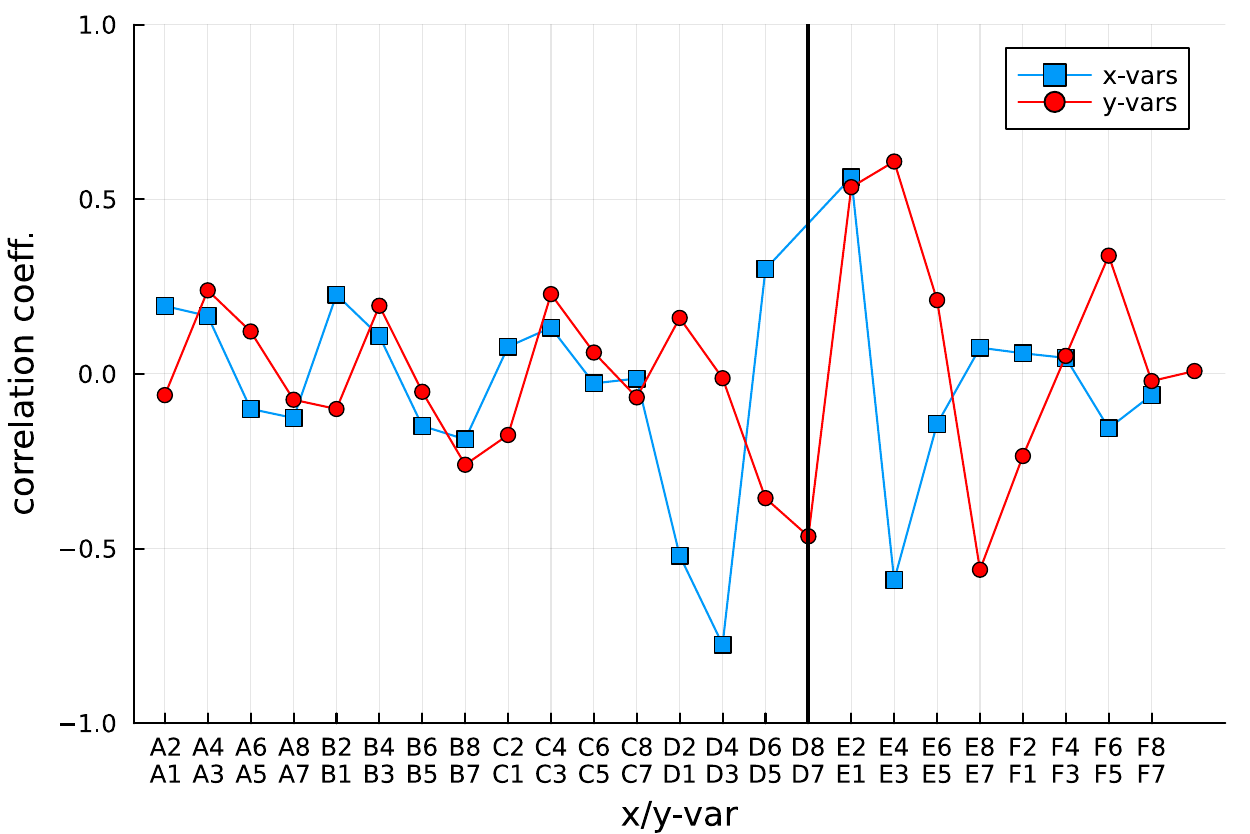}
    \includegraphics[width=0.7\linewidth]{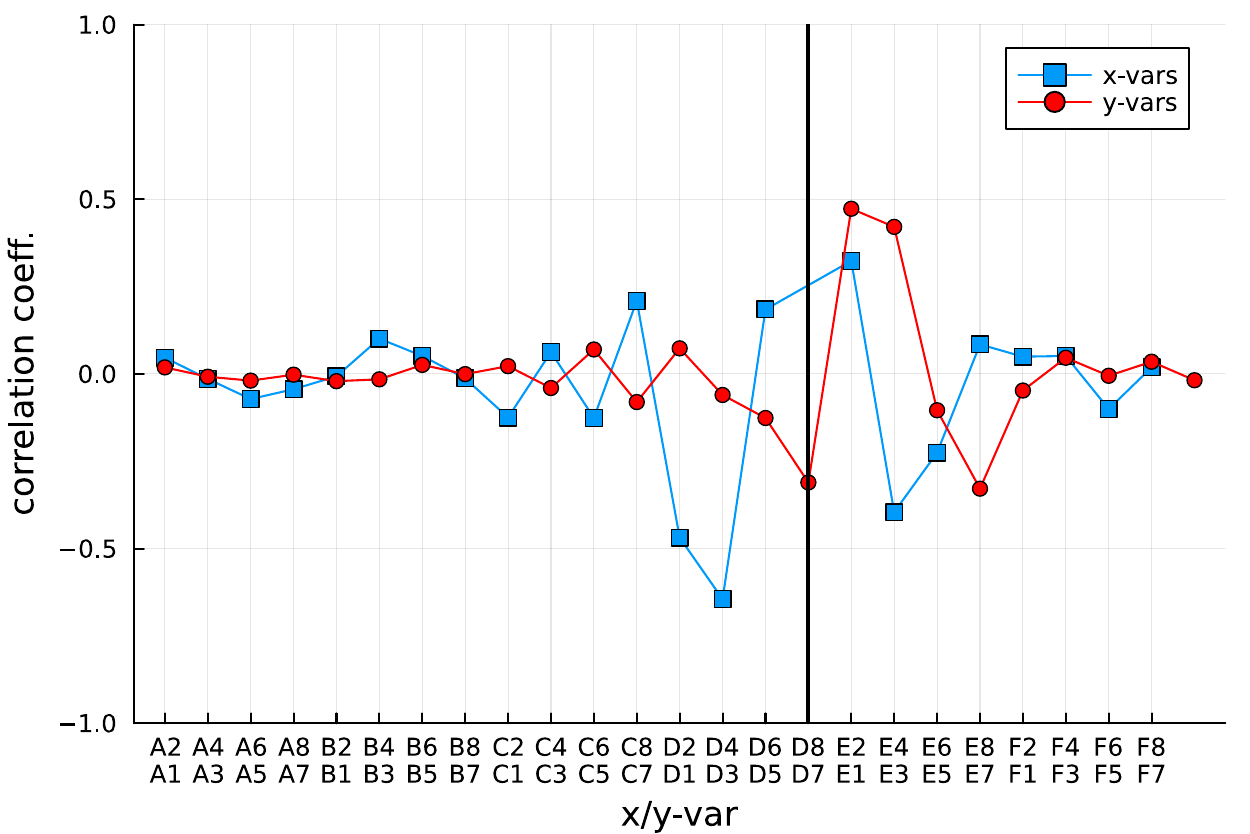}
    \caption{The correlation coefficient between var D8 and the other vars, obtained from the pooled posteriors using the $-22A\to +22A$ datasets. The upper figure shows the result from our final inference with real experimental data, and the lower figure the same result obtained from an analysis with real data replaced by simulated data with the vars set equal to our final point estimates.}
    \label{fig:D8correlation-final}
\end{figure}

The correlation matrix of the var posteriors is shown in Fig.~\ref{fig:cormat_final}, and specifically for D8 in Fig.~\ref{fig:D8correlation-final}. We observe a consistent pattern of correlations of a similar form to that seen before in Fig.~\ref{fig:D8correlation}, where the vars are strongly correlated to their nearest neighbors in the same plane and then strongly anticorrelated to their next-to-nearest neighbors; this pattern repeats for the quadrupoles in the other plane, although offset by one. An oscillatory pattern of correlations between the D8 var and other, more distant vars is also observed in this figure that was not seen in Fig.~\ref{fig:D8correlation}. This phenomenon is not unique to D8; rather, we found a similar pattern for most (but not all) other vars. To investigate this further we repeated the inference but replaced the real BPM measurements with those obtained from Bmad simulations with the vars set equal to the central values of our results in Tab.~\ref{tab-final-vars-all}. The posterior correlations between var D8 and other vars for this experiment is also shown in Fig.~\ref{fig:D8correlation-final}, where we see similar correlations with neighboring vars but little evidence of these longer distance oscillatory patterns. Similar behavior was found for the other vars. This suggests that the effect is not caused by simple beam dynamics but some other, more subtle effect.

%A similar oscillatory pattern is observed to a lesser extent between all of the vars in Fig.~\ref{fig:D8correlation-final}, and also in the pattern of the vars themselves in Fig.~\ref{fig:vars_final}. EXPLANATION SEEMS POSSIBLE GIVEN DISCUSSION WITH DOMAIN SCIENTISTS

%x-var and y-var curves are 180 degrees out of phase in Fig.~\ref{fig:D8correlation} and Fig.~\ref{fig:D8correlation-final} because quadrupole focuses in one plane and defocuses in the other. The sinusoidal shape could be a result of the betatron tune of the Booster (between 4 and 5). TODO: Chris will use fake data to redo the analysis to see whether we get a cleaner sinusoidal pattern.

%POSSIBLY THE OFFSET BY 1 REFLECTS THE DIRECTION THE BEAM IS FLOWING IN THE RING? 
%HOW CAN WE EXPLAIN THE NEIGHBOR CORRELATIONS GIVEN THAT THE MAGNET LAYOUT IS   D QH D QV D QH D QV D QH D QV? IS THE OSCILLATORY PATTERN DUE TO THE OSCILLATING NATURE OF THE BEAM ITSELF; DOES THE WAVELENGTH CORRESPOND IN SOME WAY TO THE WAVELENGTH OF BEAM FLUCTUATIONS?

\begin{figure}[!htb]
    \centering
    \includegraphics[width=0.48\columnwidth]{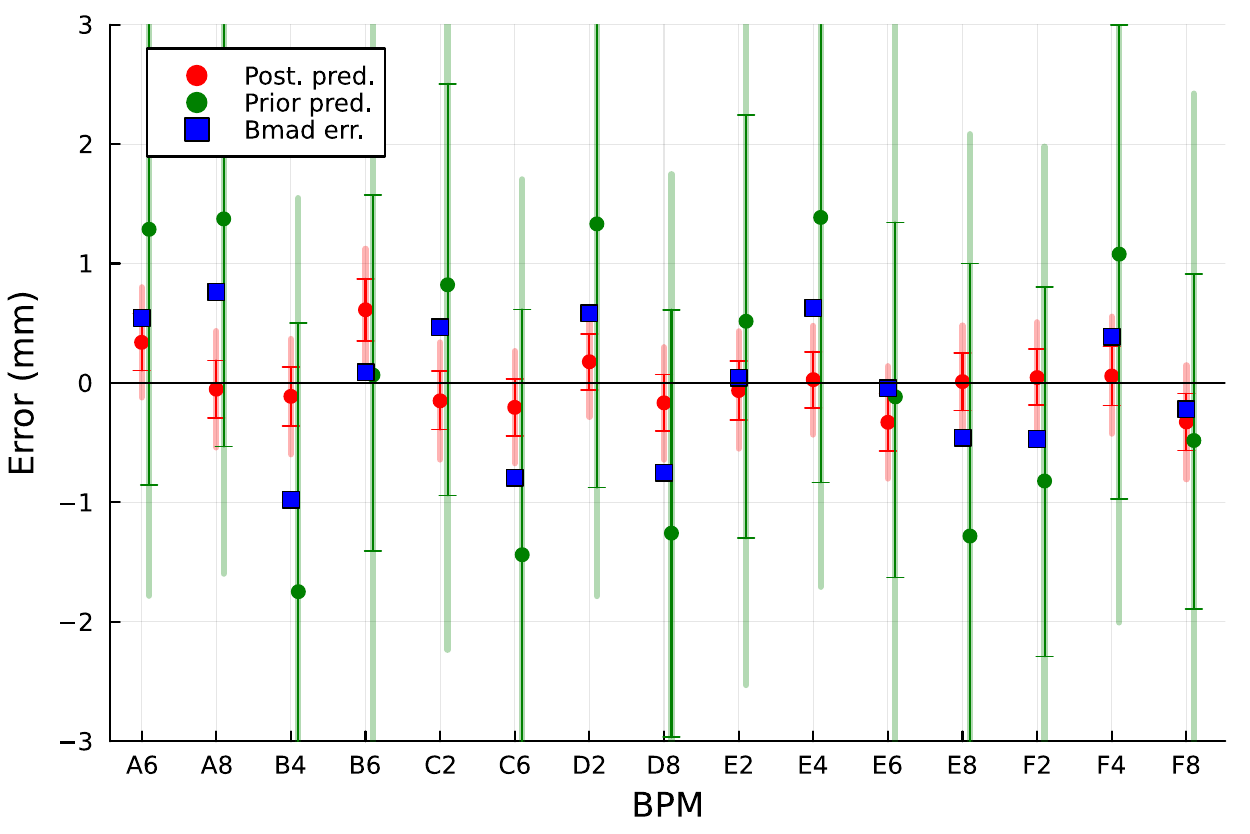} 
    \includegraphics[width=0.48\columnwidth]{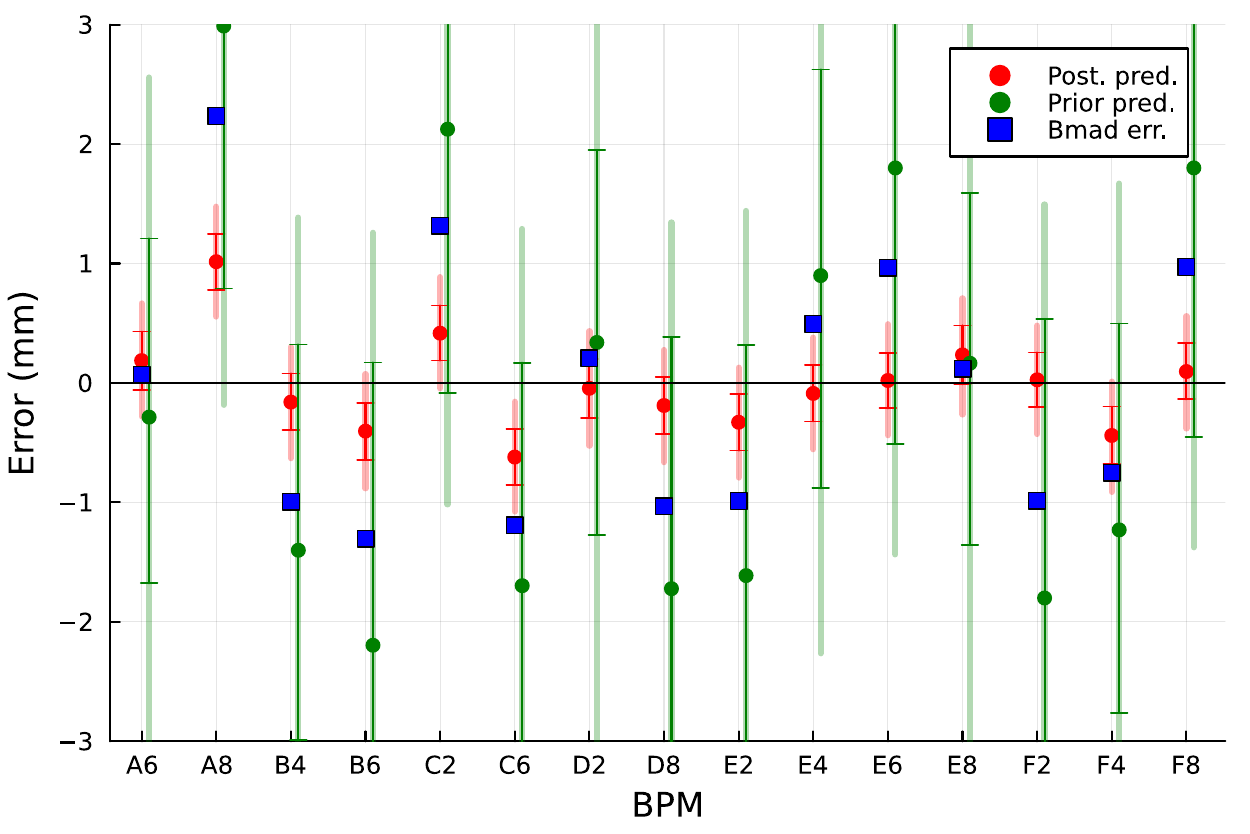} 
    \caption{Examples of the absolute deviation between the data and the prior and posterior predictions, and also between the data and the Bmad digital twin, as a function of the BPM. The dark and light error bars give the 68\% and 95\% confidence intervals, respectively. These results were obtained for the first two horizontal plane observations in the first $-22A \to +22A$ dataset.
    }
\label{fig:post_prior_pred_x}
\end{figure}

\begin{figure}[!htb]
    \centering
    \includegraphics[width=0.7\columnwidth]{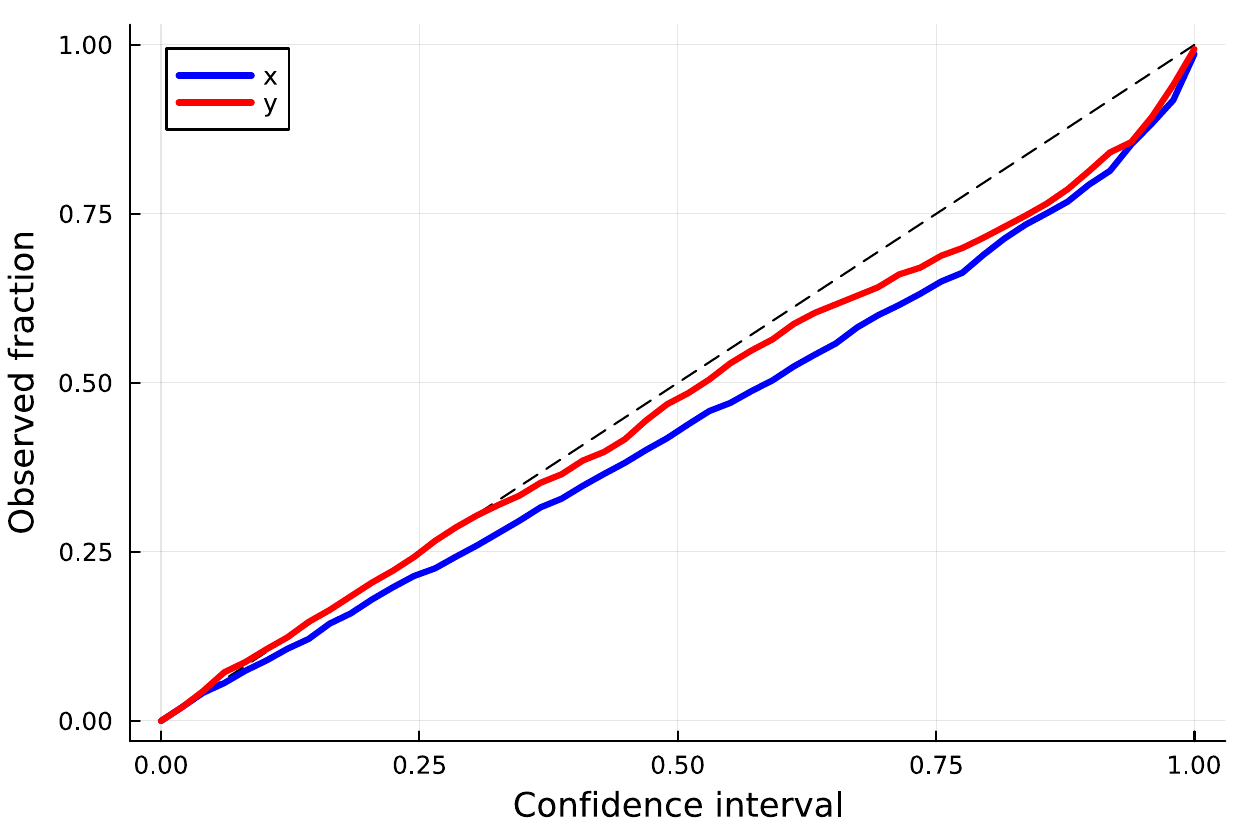} 
    \caption{The calibration curves for the $x$ (blue) and $y$ (green) orbit measurements, taken over all BPMs and all four $-22A \to +22A$ datasets. The horizontal axis gives the confidence interval and the vertical axis the observed fraction of measurements within this interval of the posterior predictive distribution. The dashed diagonal line gives the expected result.
    }
\label{fig:calib_overall}
\end{figure}

\begin{figure*}[tb]
    \centering
    \includegraphics[width=0.24\textwidth]{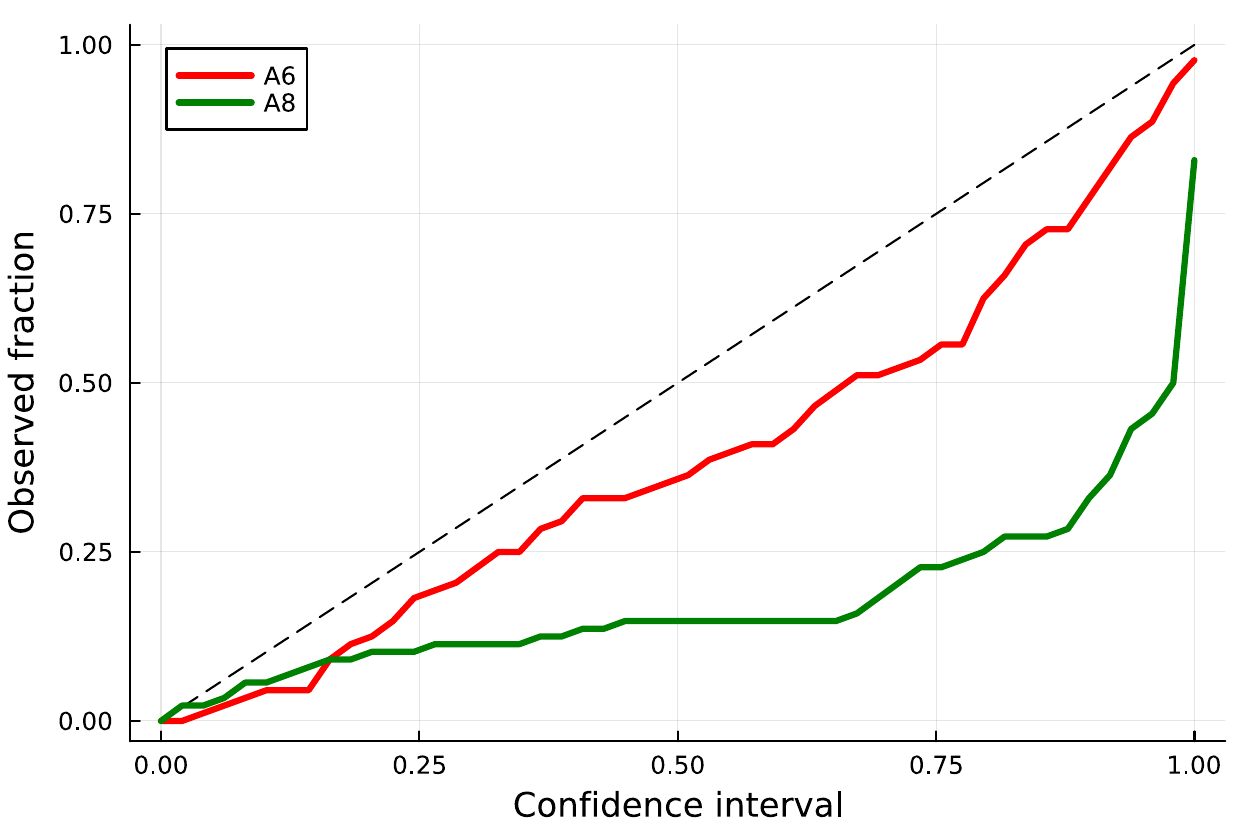}
    \includegraphics[width=0.24\textwidth]{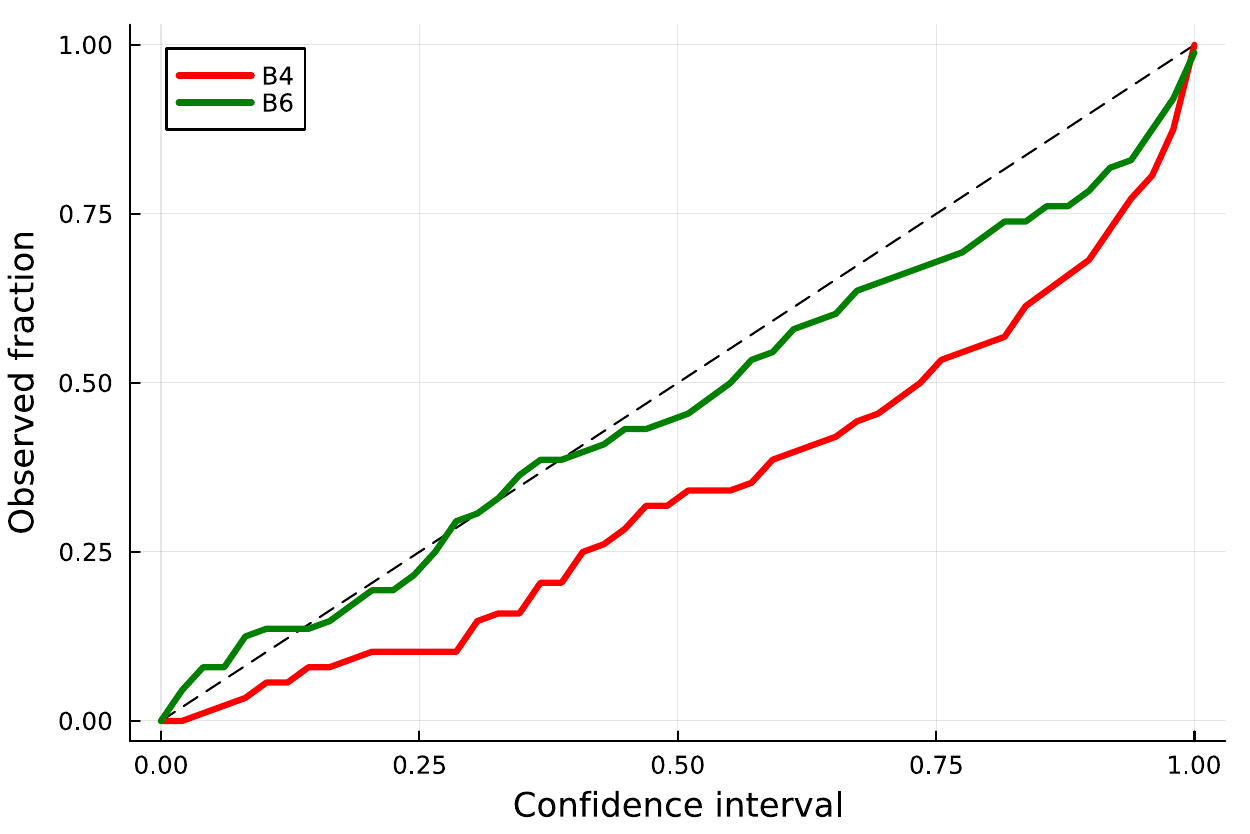}
    \includegraphics[width=0.24\textwidth]{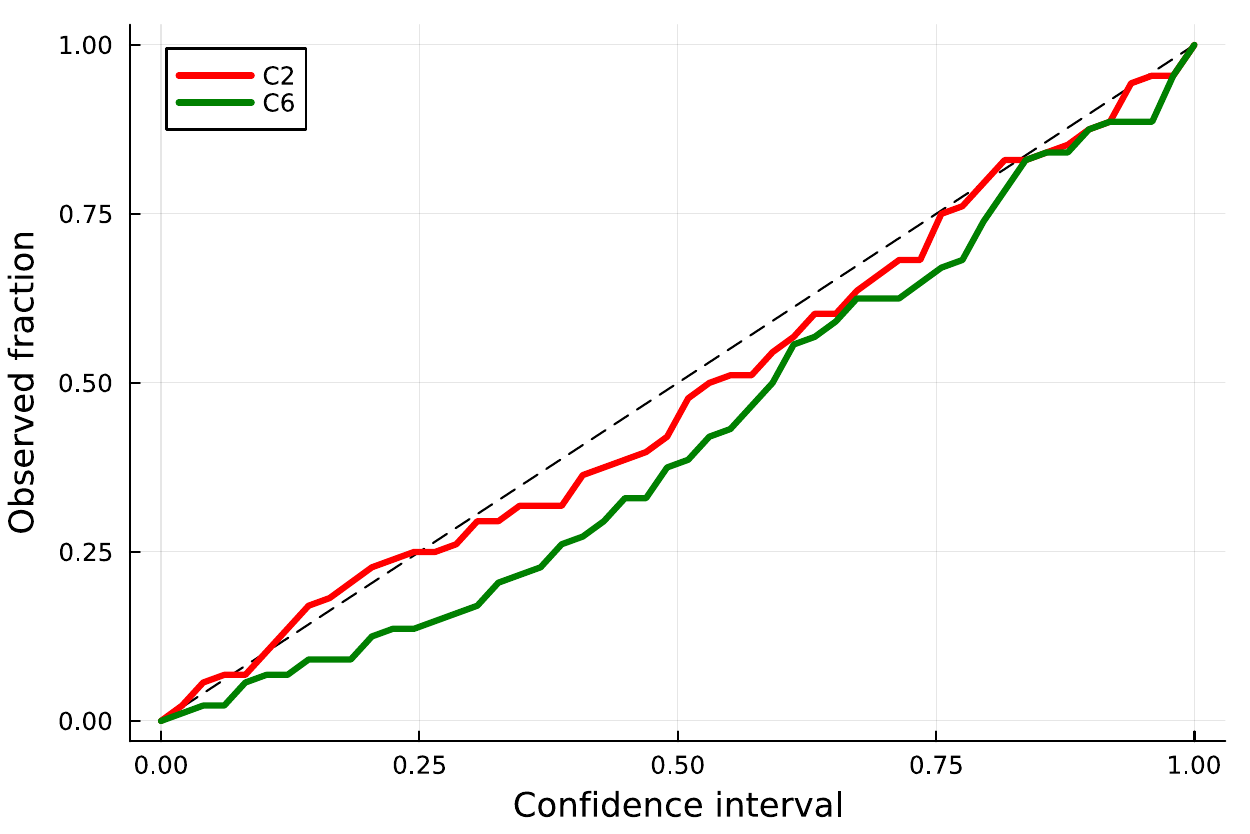}
    \includegraphics[width=0.24\textwidth]{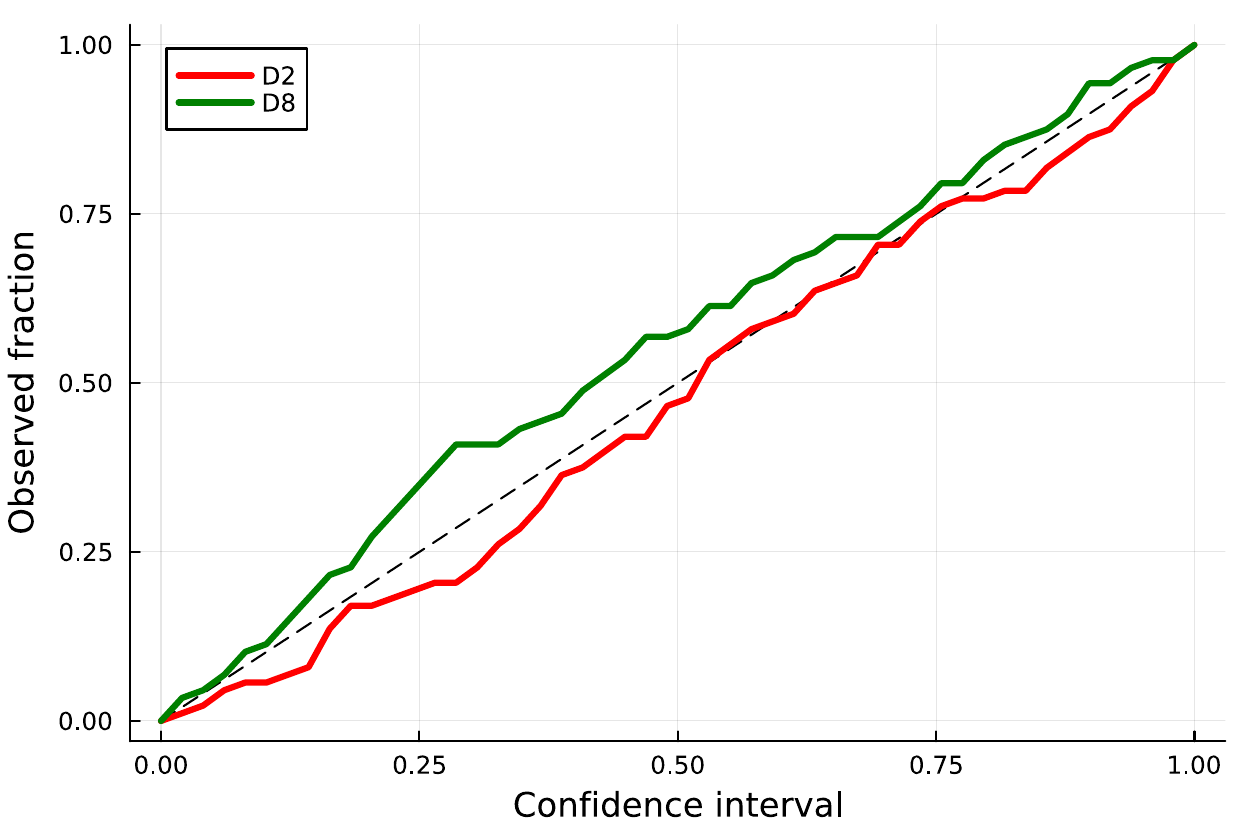}
    \includegraphics[width=0.24\textwidth]{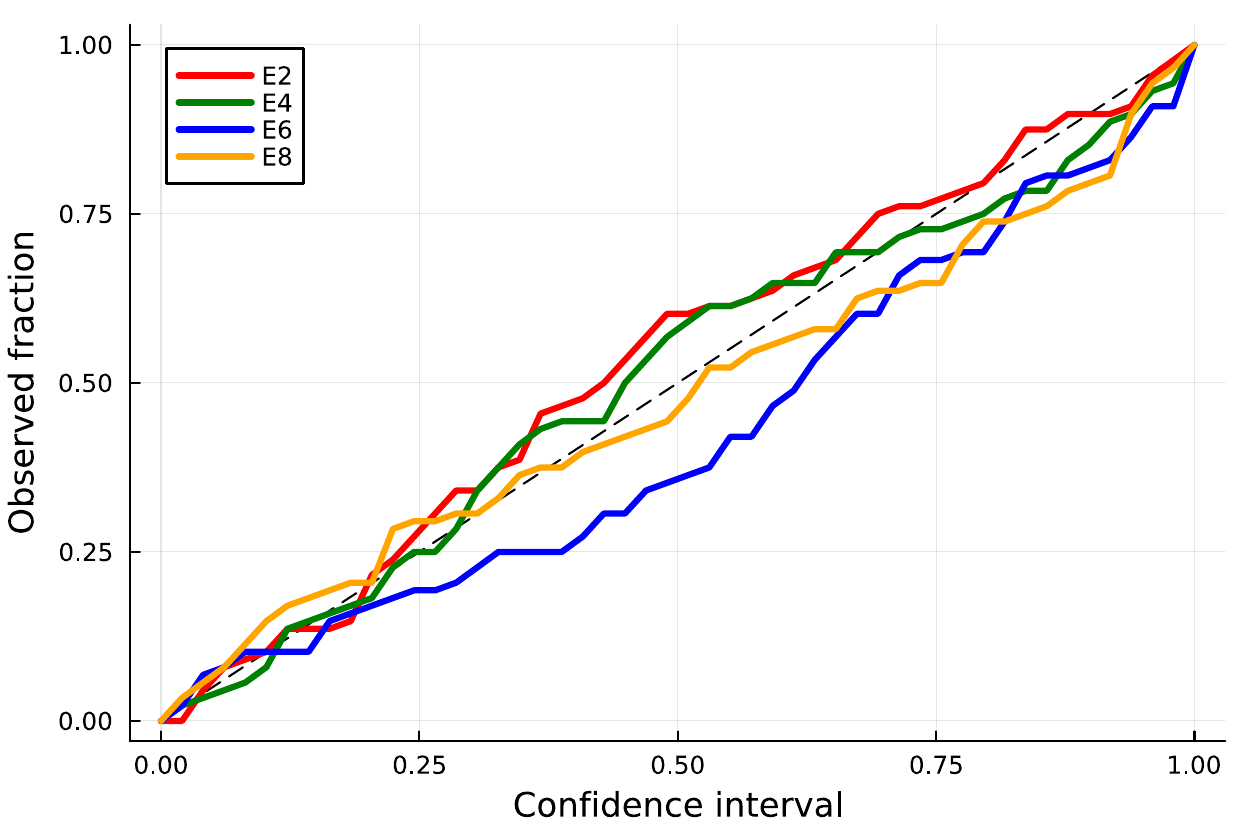}
    \includegraphics[width=0.24\textwidth]{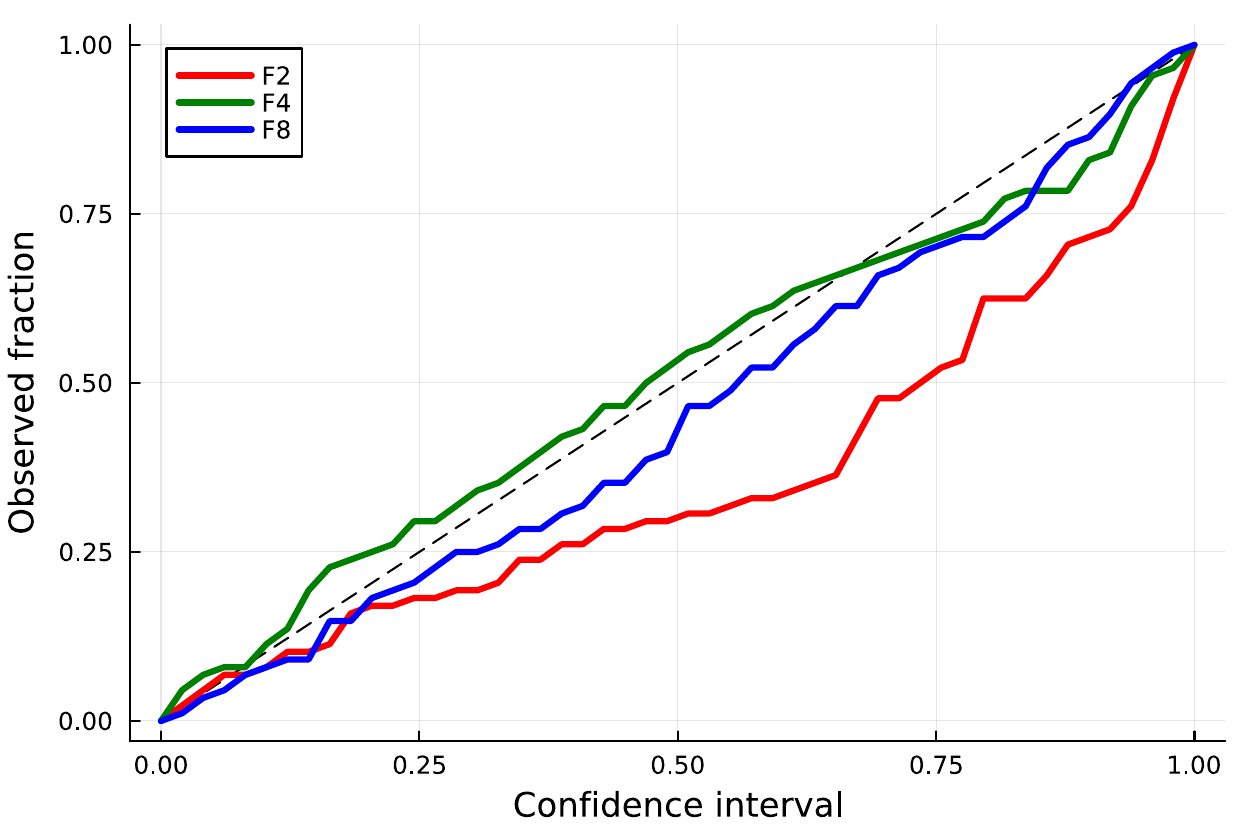}\\
    \includegraphics[width=0.24\textwidth]{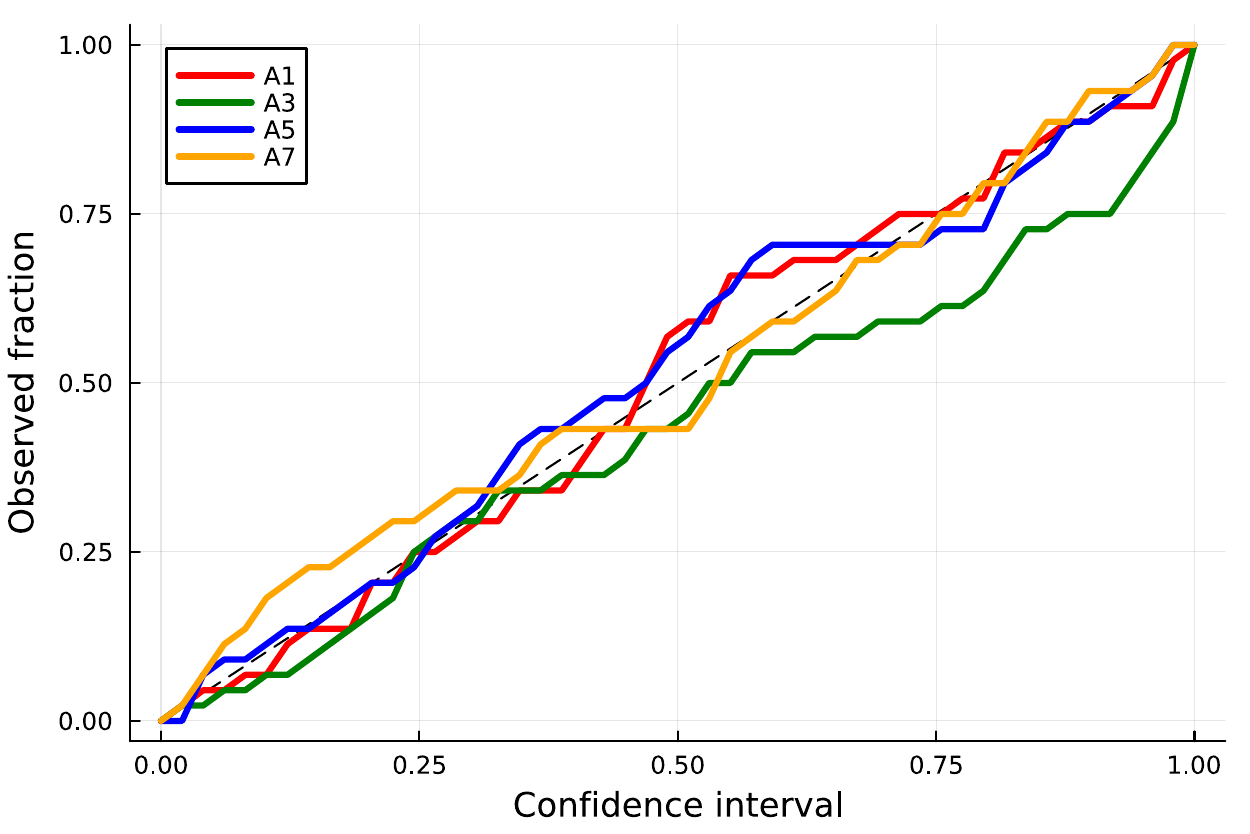}
    \includegraphics[width=0.24\textwidth]{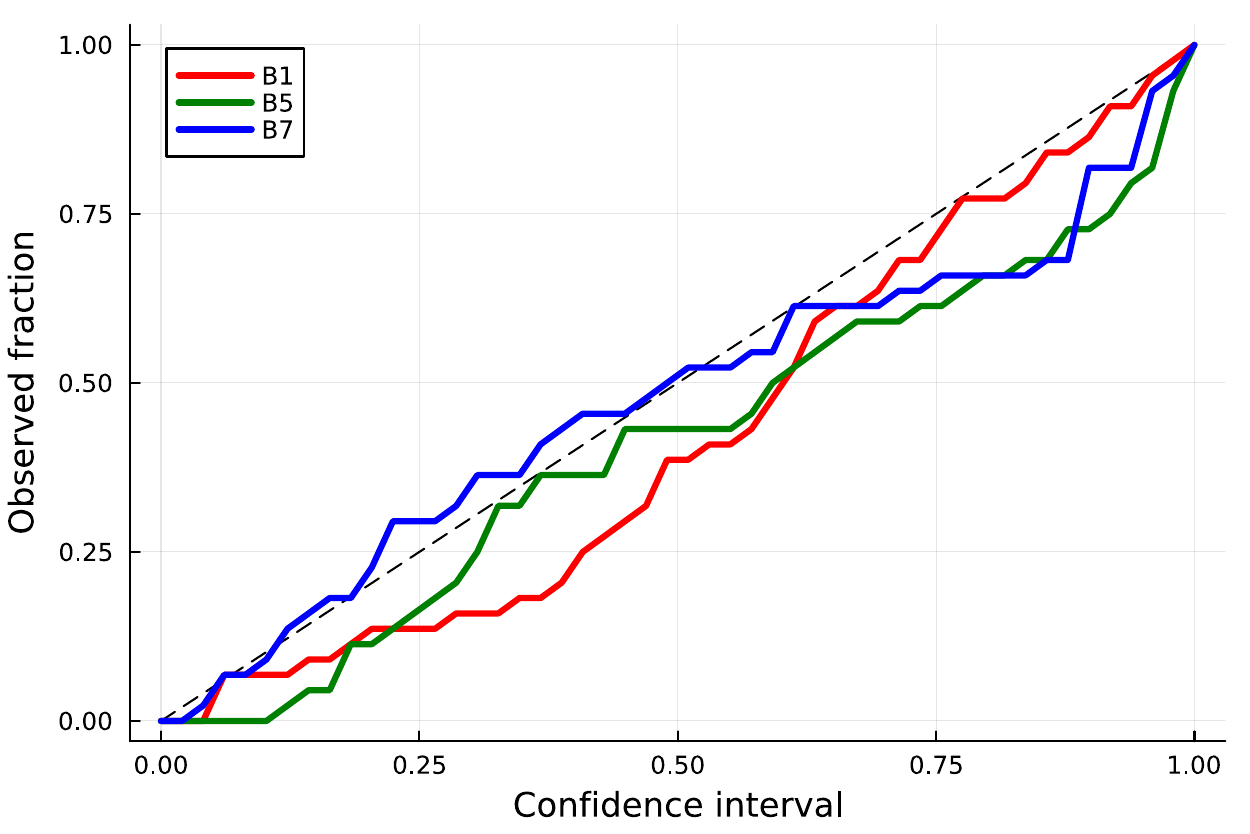}
    \includegraphics[width=0.24\textwidth]{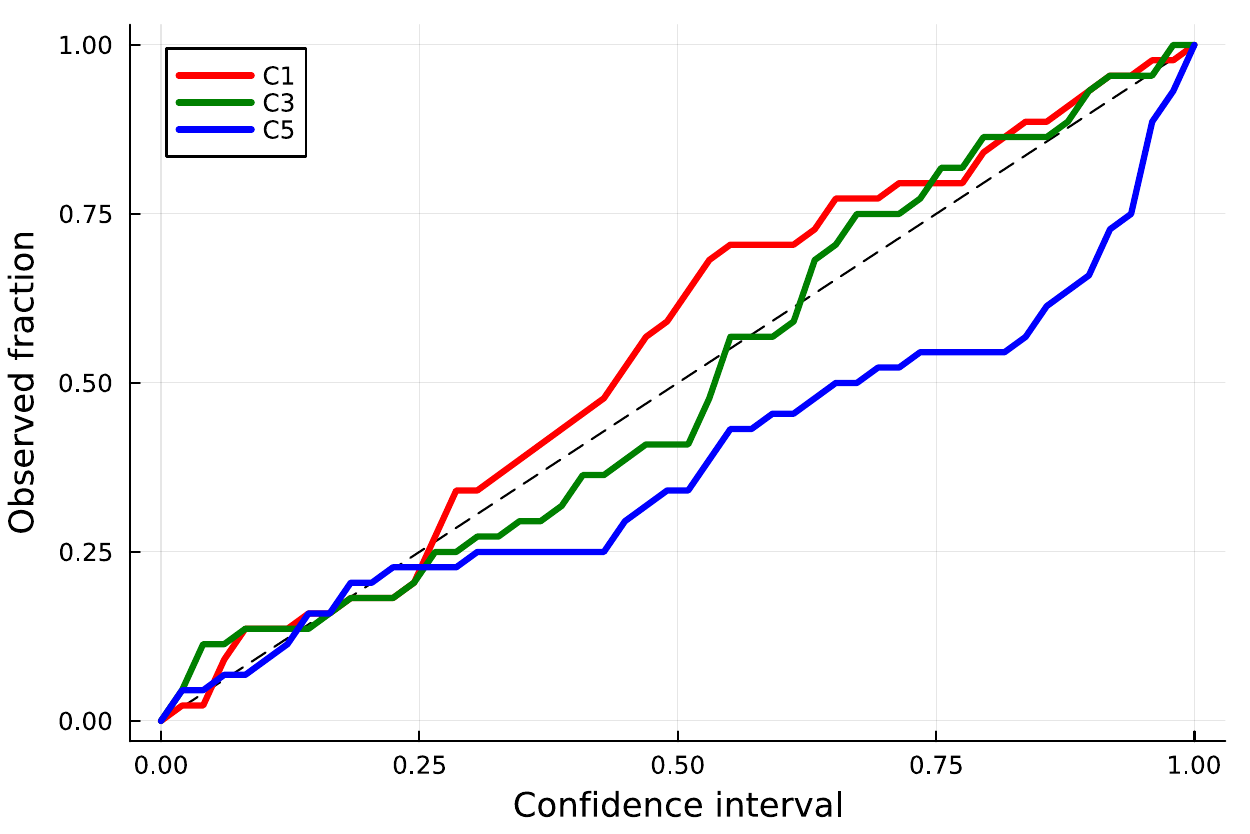}
    \includegraphics[width=0.24\textwidth]{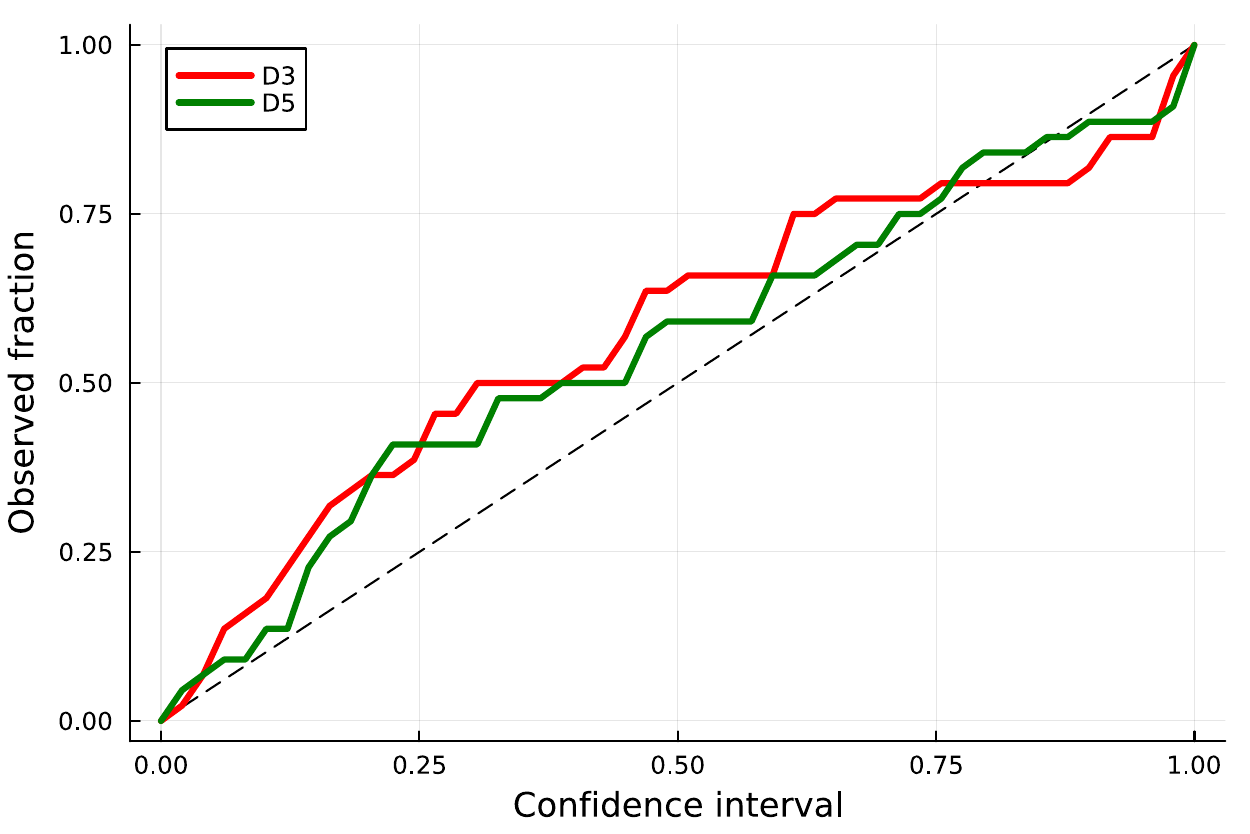}
    \includegraphics[width=0.24\textwidth]{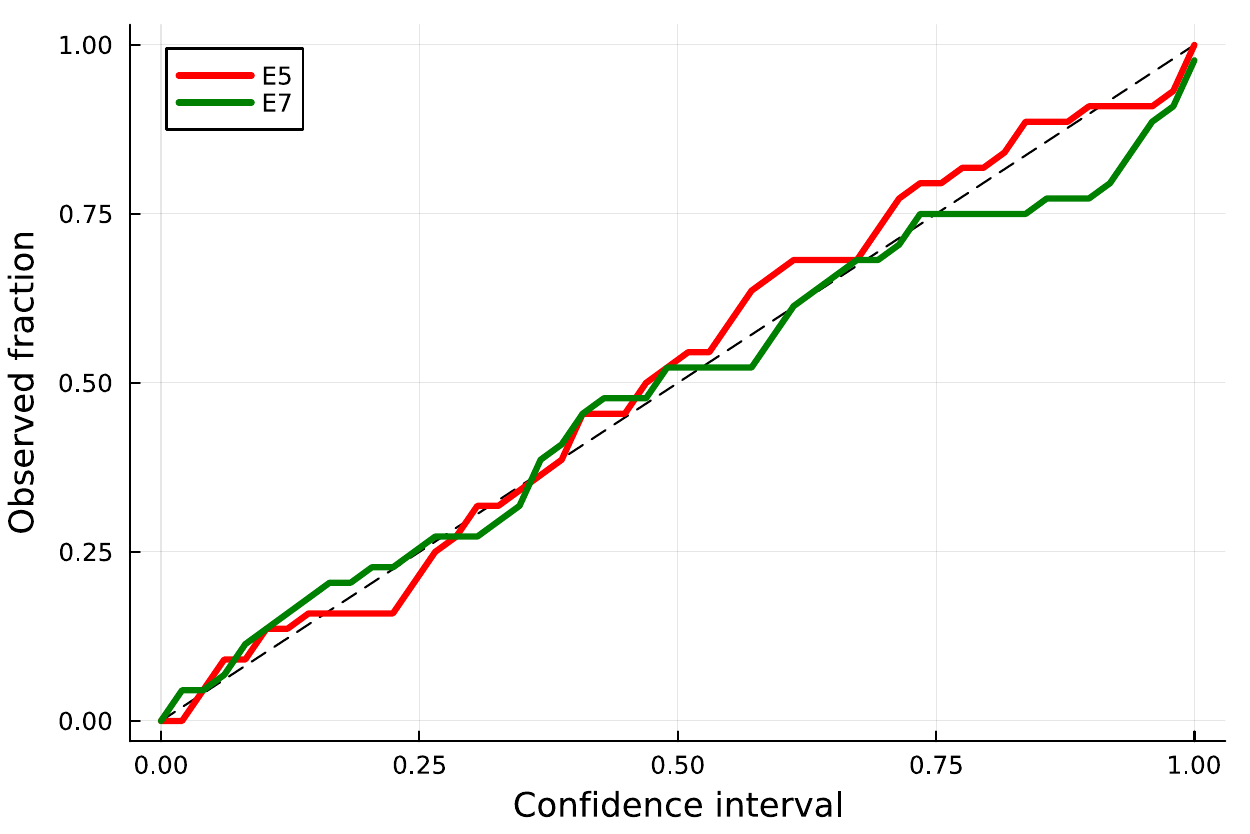}
    \includegraphics[width=0.24\textwidth]{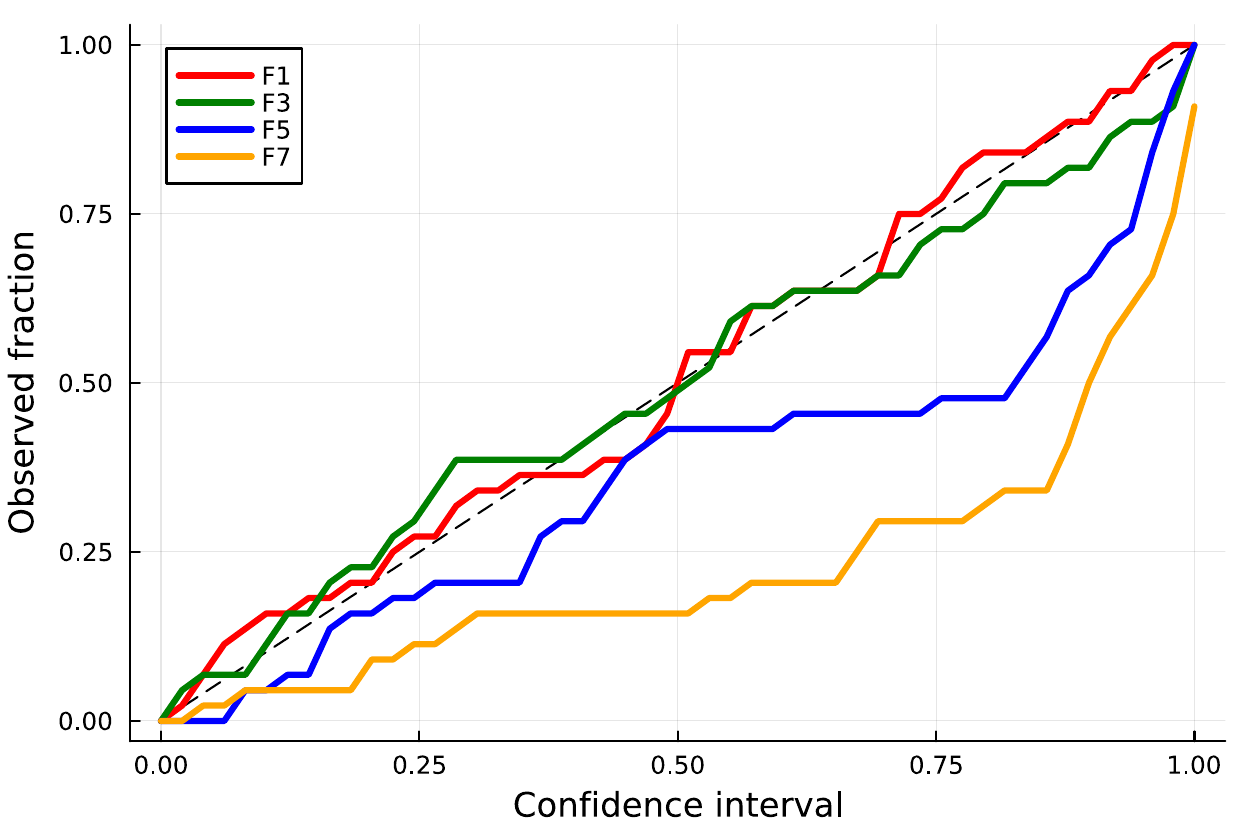}\\
    \caption{The calibration curves for the individual $x$ (upper two rows) and $y$ (lower two rows) BPMs, grouped by their associated ring segment. The horizontal axis gives the confidence interval and the vertical axis the observed fraction of measurements within this interval of the posterior predictive distribution. The dashed diagonal line gives the expected result.
    }
\label{fig:calib_segment}
\end{figure*}

Combining random resampling of the posterior (prior) distribution of the vars with the likelihood function evaluated at each of the current settings in the four $-22A \to +22A$ datasets provides the corresponding posterior (prior) predictive distributions. Examples of the absolute deviation between the data and the posterior/prior predictive results are given in Fig.~\ref{fig:post_prior_pred_x}. We observe that the range of prior predictions is very large, owing to the prior uncertainty on the vars. Conditioning on the data results in much smaller uncertainties and results that generally agree better with the data than the original Bmad predictions (also shown), indicating that the parameters successfully account for much of the deviation between the digital twin and the data. We also note that some of the deviations between the posterior predictions and the data fall outside of the 95\% confidence bands. This may indicate underestimation of the uncertainties, or that the parameters do not have sufficient power to parametrize all of the errors in the simulation; it may also simply be due to statistics. To address this we consider the calibration curves, Fig.~\ref{fig:calib_overall}, which plot the measured fraction of data points within a given confidence interval about their associated posterior predictive distributions, taken over all BPMs and all four $-22A \to +22A$ datasets. For example, in the perfect case, 68\% of data points would lie within the 68\% confidence band. We observe that the distribution of data about the posterior prediction is generally well described by our uncertainties, with only a little overconfidence (underestimation of errors) observed for larger confidence intervals, suggesting the tails of the distributions are less well described. In Fig.~\ref{fig:calib_segment} we plot separate calibration curves for each BPM, where we observe similar behavior with the notable exceptions of BPMS A8 and F7, for which we find significant overconfidence. This underestimation of the uncertainty on BPM A8 likely accounts for the large outside-of-error deviation between the posterior prediction and the data in the right plot of Fig.~\ref{fig:post_prior_pred_x}, while other, smaller deviations appear to be due to statistical effects. We intend to further investigate this overconfidence for these specific BPMs in our future research.

\subsection{Impact of incorporating point-estimate results into the digital twin}

\begin{figure}[htbp]
    \centering
    \includegraphics[width=0.7\columnwidth]{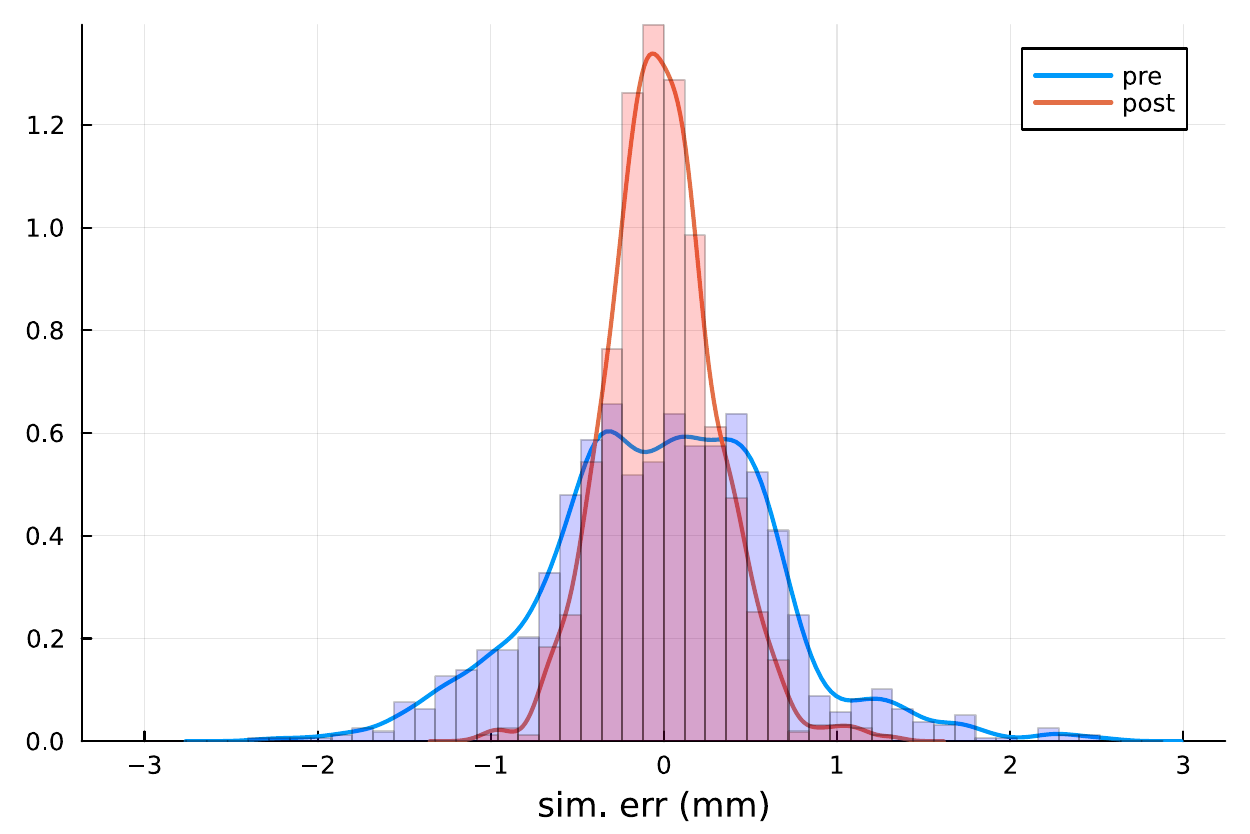} \\
    \includegraphics[width=0.7\columnwidth]{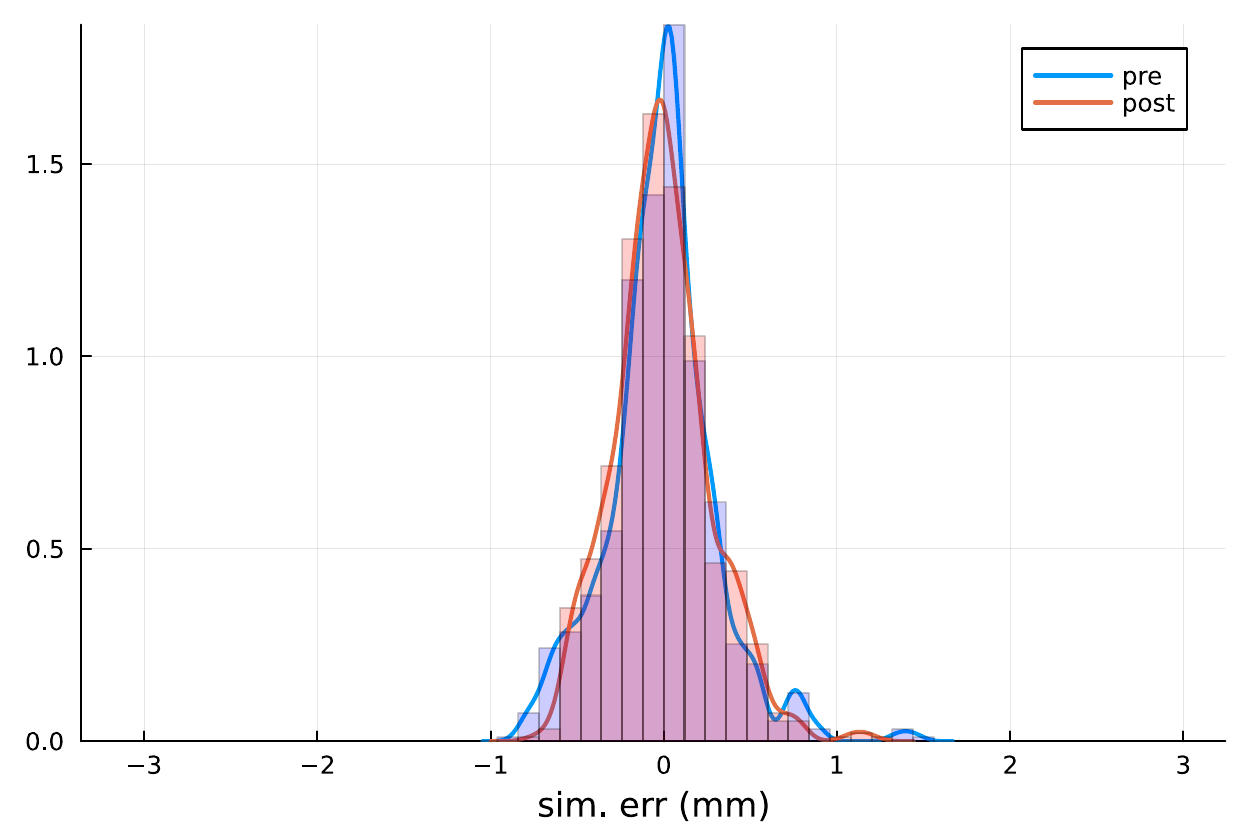} 
    \caption{The distribution of the absolute digital twin error in the x- (upper) and y-plane (lower) over all measured orbit responses within the four $-22A\to +22A$ datasets, obtained with (red) and without (blue) incorporating the point estimates of the vars.     }
\label{fig:point_est}
\end{figure}

In the posterior predictions shown above, the likelihood was evaluated used the machine-learned surrogate model rather than the true digital twin. To demonstrate whether our analysis results actually improve the digital twin we take the point estimates of the vars (the posterior mean values) from Tab.~\ref{tab-final-vars-all} and incorporate those directly into the Bmad digital twin. Note that this disregards any uncertainty on the vars. The impact upon the distribution of simulation errors over the measured orbit responses in the $-22A\to +22A$ datasets is shown in Fig.~\ref{fig:point_est}. We find that the parameters effectively decrease the simulation errors, especially in the horizontal plane where a factor of two improvement is observed. 

Note that taking the point estimates as the optimal values assumes a linear mapping between the vars and the simulated orbits. A better agreement between the digital twin and the data could be obtained were the nonlinear effects taken into account, something that we intend to investigate in the future.

\section{Conclusions}

In this work, we employed Bayesian uncertainty quantification to identify the effect of errors in simulating the quadrupole magnets on the Booster orbit response. We demonstrate that incorporating the resulting error estimates can significantly improve the agreement between simulation model and real measurements, making UQ a powerful tool for accelerator digital twin development.

\section{Appendix A: Data outlier trimming}

\begin{figure}[!htb]
    \centering
    \includegraphics[width=0.75\columnwidth]{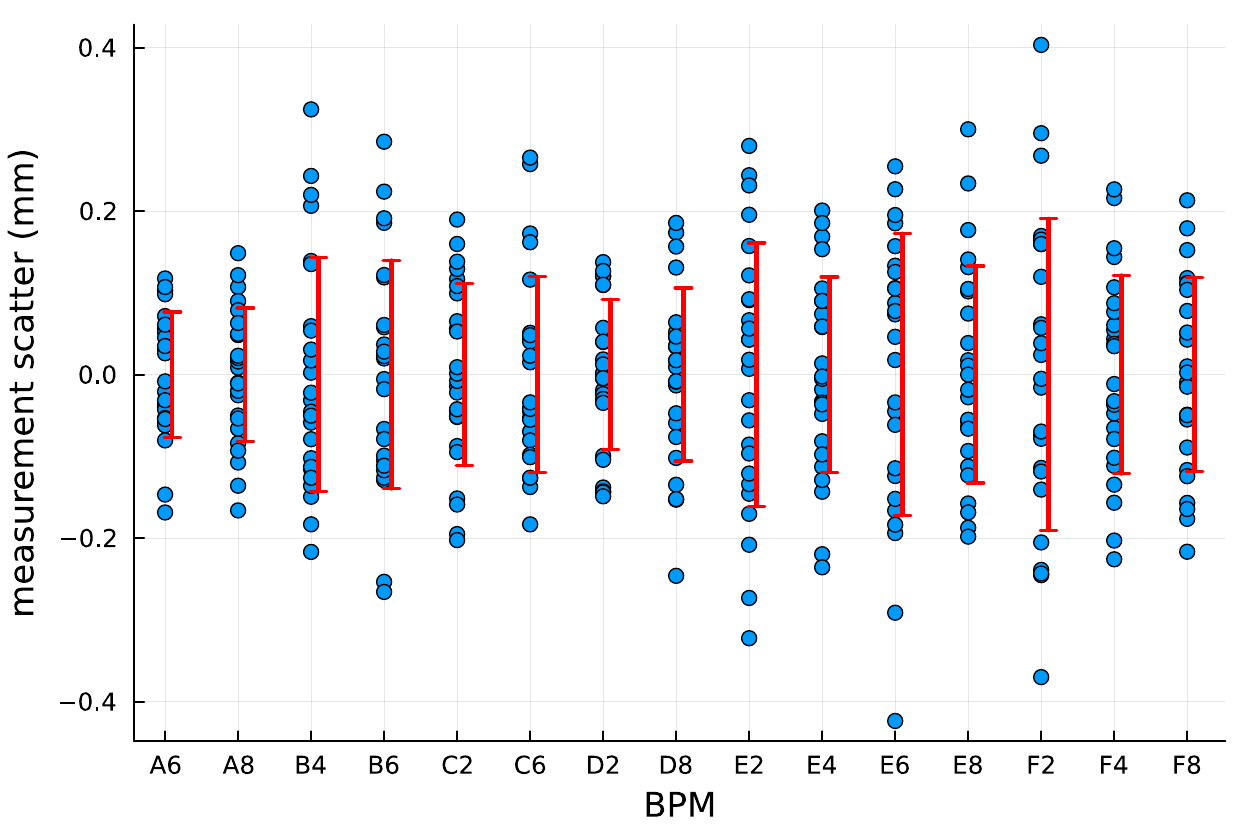} \\
    \includegraphics[width=0.75\columnwidth]{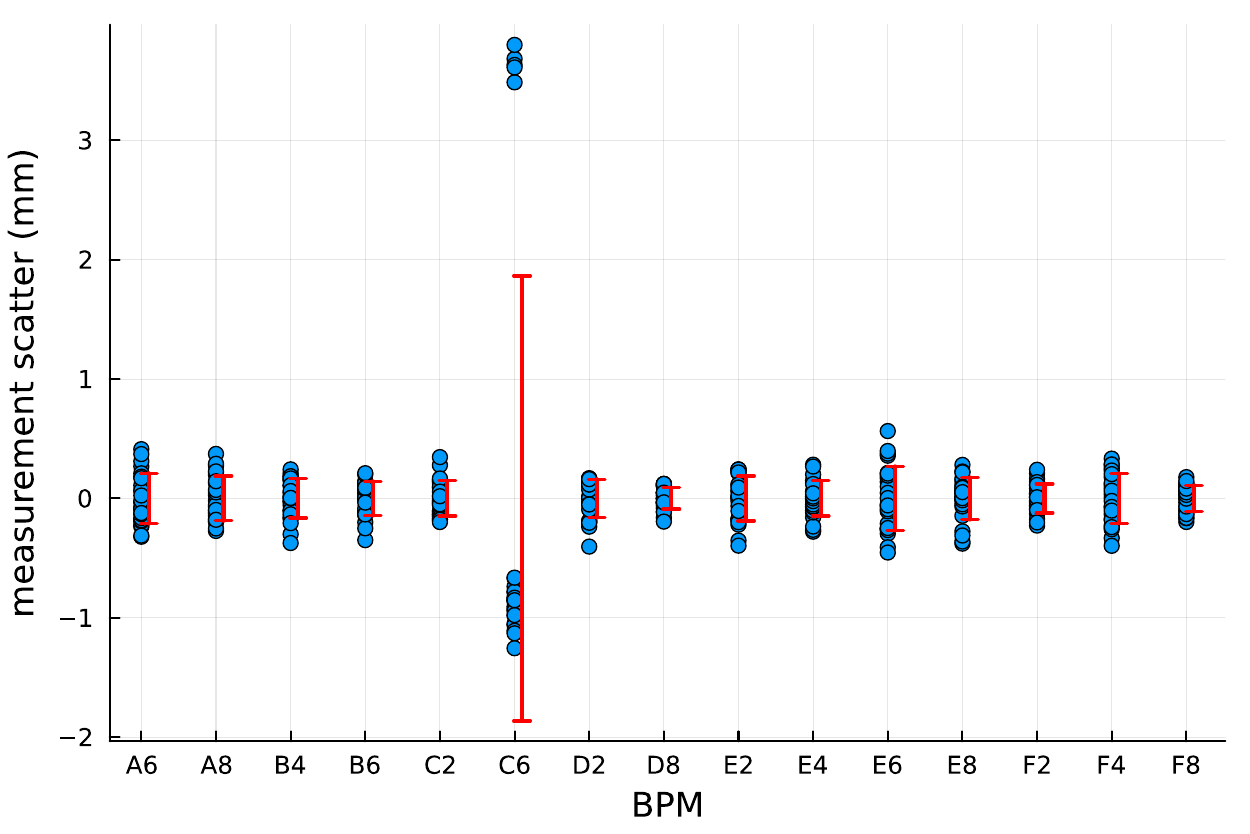}
    \caption{The scatter of orbit response measurements for perturbing corrector A2 (upper) and A6 (lower) from zero to positive kick, shown for each horizontal-plane BPM. The error bar shows the standard deviation of the points.}
    \label{fig:meas_scatter_a2_a6}
\end{figure}

%As seen in Fig.~\ref{fig:ba4_92ms},  the BPM C6 has much larger measurement fluctuations due to known hardware issues. Prior to performing the inference we trimmed these and other outlier points from the data set. 

To identify the outliers we employed the following strategy: Considering each perturbed corrector in turn, we calculate the same-plane orbit response between each possible pairing of zero-kick and positive-kick data (recall there are multiple repeat measurements for each setting). We then subtract from this the corresponding simulated response, so as to suppress the difference arising from the fluctuations in the read-back currents. Finally, we center the resulting points around zero by subtracting from each their mean value. An example for corrector A2 is shown in the upper panel of Fig.~\ref{fig:meas_scatter_a2_a6}, where the measurements are seen to fluctuate at the level of 0.1-0.2mm due to the aleatoric errors on the BPMs. No outliers are observed for this subset of the data. The corresponding result for corrector A6 is shown in the lower panel of the figure, where we observe BPM C6 shows variation far beyond the typical level owing to the hardware error mentioned above. The same process was repeated for the response between the zero-kick and negative-kick data. 

Through the above procedure we identified the data for BPMs B5, C6, and C8 as containing outliers. We then proceeded to remove the files containing these outliers from the data set, identifying them manually by locating outlier pairs sharing a common file in either the base or perturbed subset. We thus identified and removed 21 measurements for the horizontal correctors and 3 for the vertical.

%
% only for "biblatex"
%
\ifboolexpr{bool{jacowbiblatex}}%
	{\printbibliography}%
	{%
	% "biblatex" is not used, go the "manual" way
	
	%\begin{thebibliography}{99}   % Use for  10-99  references

}
%
% for use as JACoW template the inclusion of the ANNEX parts have been commented out
% to generate the complete documentation please remove the "%" of the next two commands
% 
%%%\newpage

%%%\include{annexes-A4}

\end{document}